
\magnification=\magstep1
\def\stackrel#1#2{\mathrel{\mathop{#2}\limits^{#1}}}
\voffset=1\baselineskip
\font\eightrm=cmr8
\font\eightbf=cmbx8
\font\eightsl=cmsl8
\font\twelvebf=cmbx12

\def\lg{{{\lower1.5ex\hbox{$<$}}\atop{\raise1.5ex\hbox{$>$}}}}
\def\truesupset{{\lower5pt\hbox{$\scriptstyle\supset$}\atop
 \raise5pt\hbox{$\scriptscriptstyle\not=$}}}
\def\truesubset{{\lower5pt\hbox{$\scriptstyle\subset$}\atop
 \raise5pt\hbox{$\scriptscriptstyle\not=$}}}
\def\E{\hbox{End}\,}
\def\End{\hbox{End}\,}
\def\Aut{\hbox{Aut}\,}
\def\Hom{\hbox{Hom}\,}
\def\Iso{\hbox{Iso}\,}
\def\Int{\hbox{Int}\,}
\def\ev{\hbox{ev}}
\def\coev{\hbox{coev}}
\def\Ev{\hbox{Ev}}
\def\Coev{\hbox{Coev}}
\def\id{\hbox{id}\,}
\def\idA{\hbox{id}_{\A}}
\def\idG{\hbox{id}_{\G}}
\def\Del{\Delta_{\D}}
\def\Delop{\Delta_{\D}^{op}}
\def\reven{\rho_{2i,2i+1}}
\def\rodd{\rho_{2i-1,2i}}
\def\Ad{\hbox{Ad}\,}
\def\Amp{{\bf Amp\,}}
\def\Ampr{{\bf Amp}_{\rho}}
\def\Rep{{\bf Rep\,}}
\def\F{{\cal F}}
\def\T{{\cal T}}
\def\Feven{{\cal F}_{even}}
\def\Fodd{{\cal F}_{odd}}
\def\Fr{{\cal F}_{\rho}}
\def\B{{\cal B}}

\def\Hil{{\cal H}}

\def\G{{\cal G}}
\def\D{{\cal D}}
\def\W{{\cal W}}
\def\A{{\cal A}}
\def\OA{{\cal A}}
\def\I{{\cal I}}

\def\one{{\bf 1}}
\def\onne{{\thinmuskip=5.5mu 1\!1\thinmuskip=3mu}}
\def\o{\otimes}

\def\mur{{\buildrel\lower6pt\hbox{$\scriptstyle
         \rightarrow$}\over\mu}}
\def\mul{{\buildrel\lower6pt\hbox{$\scriptstyle
         \leftarrow$}\over\mu}}
\def\mure{\mur_{\varepsilon}}
\def\mule{\mul_{\varepsilon}}
\def\cros{\raise1.9pt\hbox{$\scriptscriptstyle
          >$}\!\raise1.5pt\hbox{$\scriptstyle\triangleleft\,$}}

\def\C{\,{\raise1.5pt\hbox{$\scriptscriptstyle |$}
        \thinmuskip=4mu \!\!C\thinmuskip=3mu}}
\def\Z{{Z\!\!\!Z}}
\def\N{{\thinmuskip = 5.5mu I\!N\thinmuskip = 3mu}}

\vskip 3truecm
\centerline{\twelvebf Quantum Chains of Hopf Algebras}
\medskip
\centerline{\twelvebf with Order--Disorder Fields}
\medskip
\centerline{\twelvebf and Quantum Double Symmetry}
\vskip 1.5truecm
\centerline{Florian Nill\footnote{$^*$}{\eightrm E-mail:
NILL@omega.physik.fu-berlin.de\hfill\break
Supported by the DFG, SFB 288 "Differentialgeometrie und
Quantenphysik"}} \bigskip
\centerline{Institute for Theoretical Physics,
FU Berlin}
\centerline{Arnimallee 14, D-1000 Berlin 33, Germany}
 \vskip 1truecm
\centerline{Korn\'el Szlach\'anyi\footnote{$^{**}$}{\eightrm E-mail:
SZLACH@rmki.kfki.hu\hfill\break
\eightrm Supported by the Hungarian Scientific Research Fund,
OTKA--1815.}}
\bigskip
\centerline{Central Research Institute for Physics}
\centerline{H-1525 Budapest 114, P.O.B. 49, Hungary}

\vskip 2truecm
{\narrower\noindent
{\eightbf Abstract: }{\eightrm\
Given a finite dimensional $C^*$-Hopf algebra $H$ and its dual
$\hat H$ we construct the infinite crossed product
$\A=\dots\cros H\cros\hat H\cros H\cros\dots$
and study its representations. $\A$ is the observable algebra
of a generalized spin model with $H$-order and $\hat H$-disorder
symmetries.
By pointing out that $\A$ possesses a certain compressibility property
we can classify
all DHR-sectors of $\A$ --- relative to some Haag dual vacuum
representation --- and prove that their symmetry is described by the
Drinfeld double $\D(H)$. Complete, irreducible, translation
covariant field algebra extensions $\F\supset\A$ are shown to be
in one-to-one correspondence with cohomology classes of 2-cocycles
$u\in\D(H)\o\D(H)$.}

\vfill
\eject

{\bf 1. Introduction and summary of the results}\medskip
\medskip
Quantum chains considered as models of $1+1$-dimensional quantum
field theory exhibit many interesting features that are either
impossible or unknown in higher ($2+1$ or $3+1$) dimensions.
These features include integrability on the one hand and the
emergence of braid group statistics and quantum symmetry on the
other hand. The present paper deals with the latter, the problem of
quantum symmetry of the superselection sectors in a wide class of
quantum chains: the Hopf spin models.

Quantum chains in which a quantum group acts are well known for
some time; for example the XXZ-chain with the action of $sl(2)_q$
[P,PS] or the lattice Kac--Moody algebras of [AFS].
For a recent paper on the general action of quantum
groups on ultralocal quantum chains see [FNW].
However the discovery that quantum symmetries are described ---
even in the simplest models --- by truncated quasi-Hopf algebras
[MS,S] may constitute an obstruction towards
such an approach because the field algebras are either
non-associative or do not obey commutation relations with
$c$-number coefficients, both properties being automatically
assumed in any decent quantum chain.

Here we stress the point of view that an unbiased
approach to reveal the quantum symmetry of a model must be based
only on the knowledge of the quantum group invariant operators
(the "observables") that obey local commutation relations. This is
the approach of algebraic quantum field theory (AQFT) [H].
The importance of the algebraic method, in
particular the DHR theory of superselection sectors [DHR],
in low dimensional QFT has been realized by many authors (see
[FRS,BMT,Fr\"oGab,F,R] and many others).

The implementation
of the DHR theory to quantum chains has been carried out at first
for the case of $G$-spin models in [SzV]. These models have an
order-disorder type of quantum symmetry given by the double
$\D(G)$ of a finite group $G$ which generalizes the $Z(2)\times
Z(2)$ symmetry of the lattice Ising model. Since the disorder part
of the double (i.e. the function algebra $\C(G)$) is always
Abelian, $G$-spin models cannot be selfdual in the Kramers-Wannier
sense, unless the group is Abelian. Non-Abelian Kramers-Wannier
duality can therefore be expected only in a larger class of
models.

Here we shall investigate the following generalization of
$G$-spin models. On each lattice site there is a copy of a finite
dimensional $C^*$-Hopf algebra $H$ and on each link there is a
copy of its dual $\hat H$. Non-trivial commutation relations are
postulated only between neighbor links and sites where $H$ and
$\hat H$ act on each other in the "natural way", so as the link-site and the
site-link algebras to form the crossed products
("Weyl algebras" in the terminology of [N])
${\cal W}(\hat H)\equiv\hat H\cros H$ and ${\cal W}(H)\equiv
H\cros\hat H$. The two-sided infinite crossed
product $\dots\cros H\cros\hat H\cros H\cros\hat H\cros\dots$
defines the observable algebra $\A$ of the Hopf spin model. Its
superselection sectors (more precisely those that correspond to charges
localized within a finite interval $I$, the so called DHR sectors)
can be created by localized
amplimorphisms $\mu\colon\A\to\A\o\End V$ with $V$ denoting
some finite dimensional Hilbert space. The category of amplimorphisms
$\Amp\A$ plays the same role in locally finite dimensional theories
as the category ${\bf End\,}\A$ of endomorphisms in continuum theories.
The symmetry of the superselection sectors can be revealed by finding
the ``quantum group'' $\G$ the representation category of which is equivalent
to $\Amp\A$. In our model we find that $\G$ is
the Drinfeld double (also called the quantum double) $\D(H)$
of $H$.

Finding all endomorphisms or all amplimorphisms of a given observable
algebra $\A$ can be a very difficult problem in general. In the Hopf spin
model $\A$ possesses a property we call {\it compressibility},
which allows to do so. Namely if $\mu$ is an amplimorphism creating
some charge on an arbirary large but finite interval then there
exists an amplimorphism $\nu$ creating the same charge (i.e. $\nu$ is
equivalent to $\mu$, written $\nu\sim\mu$) but within an interval $I$
of length 2 (i.e. $I$ consists of a neighbouring site--link pair).
Therefore the problem of finding {\it all}
DHR-sectors of the Hopf spin model is reduced to a finite
dimensional problem, namely to find all amplimorphisms localized
within an interval of length 2. In this way we have proven that
all DHR-sectors of $\A$ can be classified by representations of
the Drinfeld double.

An important role in this reconstruction
is played by the so-called {\it universal} amplimorphisms in $\Amp\A$.
 These are amplimorphisms
$\rho\colon\A\to\A\o\D(H)$ such that for any object $\mu$ in $\Amp\A$
there exists an --- up to equivalence unique --- representation
$\beta_{\mu}$ of the double $\D(H)$ such that
$\mu\sim(\idA\o\beta_{\mu})\circ\rho$. Moreover,
we find that universal amplimorphisms $\rho$
can be chosen in such a way that they provide {\it coactions}
of $\D(H)$ on $\A$, i.e. they satisfy the equations
$$
\eqalignno{
(\rho\o\id)\circ\rho\ &=\ (\idA\o\Delta_\D)\circ\rho&(1.1a)\cr
(\idA\o\varepsilon_\D)\circ\rho\ &=\ \idA&(1.1b)\cr}
$$
where $\Delta_\D:\D(H)\to\D(H)\o\D(H)$ is the coproduct and
$\varepsilon_\D:\D(H)\to\C$ is the counit
on $\D(H)$. The quasitriangular
$R$-matrix can be determined by computing the
statistics operator of $\rho$ giving
$$
\epsilon(\rho,\rho)\ =\ \onne\o P^{12}R
  \quad\in\A\o\D(H)\o\D(H)\eqno(1.2)
$$
where $P^{12}$ is the usual transposition. The antipode
$S_{\D}$ can be recovered by studying conjugate objects $\bar\rho$
and intertwiners $\rho\times\bar\rho\to\idA$.
The statistical dimensions $d_r$ of the irreducible components
$\rho_r$ of $\rho$ are integers: they coincide with the dimensions
of the corresponding irreducible representation $D_r$ of $\D(H)$.
The statistics phases can be obtained from the universal balancing
element $s=S_{\D}(R_2)R_1\in\hbox{Center}\D(H)$ evaluated in
the representations $D_r$.

We emphasize that having
established the equivalence $\Amp\A\cong\Rep\D(H)$ does
not mean that the double $\D(H)$
could be reconstructed from the
observable algebra as a {\it unique} Hopf algebra.
Only as a $C^*$-algebra (together with a distinguished 1-dimensional
representation $\varepsilon$) it is uniquely
determined.
However the quasitriangular Hopf algebra structure on
$\D(H)$ can be recovered only up to a twisting by a 2-cocycle:
If $u\in\G\o\G$ is a 2-cocycle, i.e. a unitary satisfying
$$\eqalign{
(u\o\one)\cdot(\Del\o\id)(u)\ &=\ (\one\o
u)\cdot(\id\o\Del)(u)\,,\cr
(\varepsilon_\D\o\id)(u)\ &=\ (\id\o\varepsilon_\D)(u)\ =\ \one\cr}
\eqno(1.3)$$
then the twisted double with data
$$\eqalign{
\Delta'\ &=\ \Ad u\circ\Delta\cr
\varepsilon'\ &=\ \varepsilon\cr
S'\ &=\ \Ad q\circ S\qquad q:=u_1S(u_2)\cr
R'\ &=\ u^{op}Ru^*\cr
}$$
is as good for a symmetry as the original one. In fact, we prove in
Section 4 that for all 2-cocycles $u$ there is a universal coaction
$\rho'$ satisfying (1.1) with $\Delta'$ instead of
$\Del$. Vice versa, any universal coaction $\rho'$ is
equivalent to a fixed one
by an isometric intertwiner $U\in\A\o\G,\
(\idA\o\varepsilon_\D)(U)=\onne,\
U\rho(A)=\rho'(A)U,\ A\in\A,$ satisfying a {\it  twisted
cocycle condition}
$$(U\o\one)\cdot(\rho\o\idG)(U)\ =\ (\onne\o
u)\cdot(\idA\o\Del)(U)\,,\eqno(1.4)$$
implying the identities (1.3) for $u$.
We point out that (1.4) is a generalization of the usual notion
of cocycle equivalence for coactions where one requires
$u=\one\o\one$ [NaTa,BaSk,E].

This type of
reconstruction of the quasitriangular Hopf algebra $\D(H)$ is
a special case of the generalized Tannaka-Krein theorem
[U,Maj1]. Namely, any faithful functor
$F\colon{\cal C}\to Vec$ of strict monoidal braided rigid
$C^*$-categories to the category of finite dimensional
vector spaces factorizes as $F=f\circ\Phi$ to the forgetful
functor $f$ and to an equivalence $\Phi$ of ${\cal C}$ to the
representation category $\Rep\G$ of a quasitriangular $C^*$-Hopf
algebra $\G$. In our case ${\cal C}$ is the category $\Amp\A$ of
amplimorphisms of the observable algebra $\A$. The functor $F$ to
the vector spaces is given naturally by associating to the
amplimorphism $\mu\colon\A\to\A\o\End V$ the vector space $V$.
Although the vector spaces $V$ cannot be seen looking at only the
abstract category $\Amp\A$, they are "inherently" determined by the
amplimorphisms and therefore by the observable algebra itself.
In this respect using amplimorphisms one goes somewhat beyond the
Tannaka-Krein theorem and approaches a Doplicher-Roberts [DR] type of
reconstruction.

Section 4 is devoted to the construction and
classification of field algebra extensions $\F\supset\A$. Here
a $C^*$-algebra extension $\F\supset\A$ is called a {\it complete
irreducible field algebra} over $\A$ if
(I) $\F$ has sufficiently many fields in order to write any
amplimorphism $\mu\colon\A\to\A\o\End V$ as
$\mu(A)=F(A\o\one_V)F^*$ for some unitary $F\in\F\o\End V$,
(II) the relative commutant of $\A$ in
$\F$ is trivial, $\A'\cap\F=\C\onne$, (III) there exists a conditional
expectation ${\cal E}\colon\F\to\A$  onto $\A$ with finite index,
and (IV)  $\F$ is minimal under conditions (I--III) .
For our Hopf spin model we
classify all such field algebra extensions and find that, up to
equivalence, they are in one-to-one correspondence with cohomology
classes of 2-cocycles (1.3).
All these field algebras arise as crossed products
$\F=\A\cros\hat\G$ with respect to some coaction
$(\rho,\Delta)$, where $\hat\G$ is the Hopf algebra dual to
$(\G,\Delta)$. A convenient presentation for them can be given
by a unitary {\it master field} $F\in\F\o\G$ satisfying
$$
\eqalign{
F(A\o\one)\ &=\ \rho(A)F\quad A\in\A\cr
F^{01}F^{02}\ &=\ (\id\o\Delta)(F)\cr
F^*\ &=\ (\id\o S)(F)\equiv F^{-1}\cr
}\eqno(1.5)$$
where $F^{0i}$, for $i=1,2$, denotes the obvious embedding of $F$
into $\A\o\G\o\G$.
In other words the map $\hat\G\ni\xi\to F(\xi)\equiv
(\idA\o\xi)(F)\in\F$ provides a $*$-algebra inclusion.
The fields $F(\xi)$ are precisely the generalizations of the
order-disorder fields in [SzV].

The field algebras $\F$ carry a natural action $\gamma$ of the Hopf
algebra $\G$ as a global gauge symmetry
such that $\A$ coincides with the subalgebra of
$\G$-invariant elements:
$$
\{A\in\F\,|\,\gamma_X(A)=A\varepsilon(X),\,\forall
X\in\G\}=\A\,.
$$
If $\xi=D_r^{ij}\in\hat\G$ are chosen to be the matrix elements in an
irreducible representation of $\G$ then the fields
$F_r^{ij}\equiv F(D_r^{ij})$ provide a $\G$-covariant
matrix multiplet of field operators.

Studying the problem of translation covariance of field algebras
we find that in the Hopf spin model all universal coactions are
translation covariant,
i.e. there exists an isometric intertwiner $U$ from $\rho$ to its
translate $\rho^{\alpha}$ satisfying the cocycle condition
(1.4) with $u=\one\o\one$. This guarantees that on the corresponding
field algebras
$\F$ there exists an extension of the translation automorphism
$\alpha$ commuting with the global symmetry action $\gamma$.

We also show that the master field $F$ and its translated images
$F'=(\alpha^n\o\idG)(F),\ n\in {\bf N}$, satisfy braided
commutation relations given by
$$
F'_{01}F_{02}=F_{02}F'_{01}(\onne\o R)
$$
where the quasitriangular R-matrix is given by (1.2).

Our result showing that there are as many complete irreducible
translation covariant field algebras as cohomology classes of
2-cocycles must be compared with the result of [DR] on the uniqueness
of the field algebra. This apparent discrepancy may be explained
by viewing group theory within the more general setting of Hopf algebra
theory. In the higher
dimensional situation with group symmetry one has a
preferred choice of the coproduct and the $R$-matrix, while in our case
one has to consider
all possible ones and there is no analogue of the normal commutation
relations of [DR]. This causes that we encounter a whole family of
inequivalent field algebra extensions, among which one cannot make any
``observable'' distinction.

\bigskip
{\bf 2. The structure of the observable algebra}
\bigskip
{\sl 2.1. $\A$ as an iterated crossed product}
\medskip
In this section we describe a canonical method by means of which
one
associates an observable algebra $\A$ on the 1-dimensional lattice
to any $C^*$-Hopf algebra $H$. Although a good deal of our
construction works for infinite dimensional Hopf algebras as well,
we restrict the discussion here to the finite dimensional case.

Consider $\Z$, the set of integers, as the set of cells of the
1-dimensional lattice: even integers represent lattice sites, the
odd ones represent links. Let $H=(H,\Delta,\varepsilon,S,*)$ be a
finite dimensional $C^*$-Hopf algebra (see Appendix A).
We denote by $\hat H$
the dual of $H$ which is then also a $C^*$-Hopf algebra. We denote
the structural maps of $\hat H$ also by $\Delta,\varepsilon,S$,
Elements of $H$ will be
typically denoted as $a,b,\dots$, while those of $\hat H$ by
$\varphi,\psi,\dots$. The canonical pairing between $H$ and $\hat
H$
is denoted by $a\in H, \varphi\in\hat H\mapsto\langle
a,\varphi\rangle \equiv\langle\varphi,a\rangle$.
 There are natural actions of $H$
on $\hat H$ and $\hat H$ on $H$ given by the Sweedler's arrows:
$$\eqalign{
 a\rightarrow\varphi&=\varphi_{(1)}\langle\varphi_{(2)},a\rangle\cr
 \varphi\rightarrow a&=a_{(1)}\langle
a_{(2)},\varphi\rangle\cr}\eqno(2.1)$$
Let us associate to each even integer $2i$ a copy $\A_{2i}$ of the
$C^*$-algebra $H$ and to each odd integer $2i+1$ a copy
$\A_{2i+1}$ of $\hat H$. We specify isomorphisms $a\in H\mapsto
A_{2i}(a)\in\A_{2i}$ and $\varphi\in\hat H\mapsto
A_{2i+1}(\varphi)
\in\A_{2i+1}$ such that $A_{2i}(a)\mapsto A_{2j}(a), a\in H$ and
$A_{2i+1}(\varphi)\mapsto A_{2j+1}(\varphi), \varphi\in\hat H$ are
isomorphisms for each integer $i$ and $j$.
 The algebra $\A$ of observables is --- by definition ---
generated by the 1-point localized algebras $\A_i$,
$$\A=C^*\hbox{-}\langle \A_i,\ i\in\Z\rangle\eqno(2.2)$$
and subjected to the following commutation relations: On the
one hand
$$[A,B]=0\quad A\in\A_i,\ B\in\A_j\ \hbox{\ whenever }\ |i-j|\geq
2\,.\eqno(2.3a)$$
On the other hand
neighbour algebras are required to generate the crossed
products $\A_i\cros\A_{i+1}\cong H\cros\hat H$ if $i$ is even
and $\cong \hat H\cros H$ if $i$ is odd. These crossed products
are understood with respect to the actions 2.1.
More explicitely we postulate the relations
$$\eqalign{
  A_{2i+1}(\varphi)A_{2i}(a)&=A_{2i}(a_{(1)})\langle
a_{(2)},\varphi_{(1)}\rangle A_{2i+1}(\varphi_{(2)})\cr
A_{2i}(a)A_{2i-1}(\varphi)&=A_{2i-1}(\varphi_{(1)})
\langle\varphi_{(2)},
  a_{(1)}\rangle A_{2i}(a_{(2)})\cr}\eqno(2.3b)$$
These relations allow to order any monomial in the $A_i$-s in
increasing order with respect to their location $i$. The existence
of the antipode ensures that we can invert relations 2.3b and
express everything in terms of decreasingly ordered monomials.
The above relations define what can be called the iterated crossed
product algebra
$$\A = \dots\cros H\cros\hat H\cros H\cros\hat H\dots
\eqno(2.4)$$
where the dots include a $C^*$-inductive limit procedure.
For an interval $I=\{i,i+1,\dots,i+n\}$ we denote by $\A(I)$ the
subalgebra generated by the $\A_j$ with $j\in I$. Elements of
$\A(I)$ are called the observables localized within $I$. An
important property of these algebras is that $\A(I)$ is simple for
all interval $I$ of even length.

$$\A(I)\cong M_N^{\,\o\,|I|/2},\quad I\in\I,\ \
|I|=\hbox{even},\eqno(2.5)$$
where $N=\hbox{dim}H$ and $M_N$ denotes the algebra of $N\times N$
complex matrices.
For $I$ of length 2 (2.5) follows from
the fact [N] that the 2-point algebras $\A_i\cros\A_{i+1}$
are isomorphic to $\End H$ or $\End \hat H$, respectively.

There is an other way to formulate the commutation relations 2.3b
using the multiplicative unitaries of [BaSk]. Choosing a basis
$\{b^s\}$ of $H$ and denoting by $\{\beta_s\}$ the dual basis of
$\hat H$, i.e. $\langle\beta_s,b^t\rangle=\delta_s^t$, we find
that the unitary elements
$$\eqalign{
V_{2i,2i+1}&:=\sum_s\ A_{2i}(b^s)\ \o\ A_{2i+1}(\beta_s)\cr
V_{2i-1,2i}&:=\sum_s\ A_{2i-1}(\beta_s)\ \o\ A_{2i}(b^s)\cr
}\eqno(2.6)$$
of $\A\o\A$ satisfy the relations
$$\eqalign{
V_{i,i+1}^{12}V_{i,i+1}^{23}\ &=\
V_{i,i+1}^{23}V_{i,i+1}^{13}V_{i,i+1}^{12}\cr
V_{i,i+1}^{13}V_{i-1,i}^{12}\ &=\
V_{i-1,i}^{12}V_{i,i+1}^{23}V_{i,i+1}^{13}\cr
}\eqno(2.7)$$

We also mention an other interesting property of the net $\A$
related to Jones' basic construction [J]. Let $I\subset
J\subset K$ be three intervals of length $n, n+1, n+2$,
respectively, such that either the
left or the right endpoints of all the three intervals coincide.
Then the algebras $(\A(I),\A(J),\A(K))$ form a Jones' triple, i.e.
the algebra $\A(K)$ arises as the basic construction associated to
the inclusion $\A(I)\subset\A(J)$.

\medskip
{\sl 2.2. $\A$ as a Haag dual net}
\medskip
The local commutation relations (2.3) of the observables suggests
that our Hopf spin model can be viewed in the more general
setting of algebraic quantum field theory (AQFT) as a local net.
More precisely
we will use an implementation of AQFT appropriate to study lattice
models in which the local algebras are finite dimensional.
Although we borrow the language and philosophy of AQFT, the
concrete mathematical notions we need on the lattice are quite
different from the analogue notions one uses in QFT on Minkowski
space.

Let $\I$ denote the set of non-empty finite
subintervals
of $\Z$. A net of finite dimensional $C^*$-algebras, or shortly a
{\it net} is a correspondence $I\mapsto\A(I)$ associating to each
interval $I\in\I$ a finite dimensional $C^*$-algebra together with
unital inclusions $\iota_{J,I}\colon\A(I)\to\A(J)$,
whenever $I\subset J$, such that for all $I\subset J\subset K$
one has $\iota_{K,J}\circ\iota_{J,I}=\iota_{K,I}$.

The inclusions $\iota_{J,I}$ will be suppressed and for $I\subset
J$ we will simply write $\A(I)\subset\A(J)$. If
$\Lambda$ is any (possibly infinite) subset of $\Z$
we write $\A(\Lambda)$ for the $C^*$-inductive limit of
$\A(I)$-s with $I\subset \Lambda$. Especially let $\A=\A(\Z)$.

For $\Lambda\subset\Z$ let
$\Lambda'=\{i\in\Z\,|\,\hbox{distance}(i,\Lambda)\geq 2\,\}$
which is the analogue of "spacelike complement"of $\Lambda$.
The net $\{\A(I)\}$ is called {\it local} if $I\subset J'$
implies $\A(I)\subset\A(J)'$, where $B'$ for a subalgebra $B$
of the global algebra $\A$ denotes its relative commutant within
$\A$.

The net $\{\A(I)\}$ is said to satisfy {\it (algebraic) Haag
duality} if
$$\A(I')'=\A(I)\qquad \forall I\in\I\,.\eqno(2.8)$$

The net $\{\A(I)\}$ is called {\it split} if for all $I\in\I$
there exists a $J\in\I$ such that $J\supset I$ and $\A(J)$ is
simple.

If $I=\{i,i+1,\dots,j\}$ then we write $\A_{i,j}$ for $\A(I)$
and $\A_i$ for $\A_{i,i}$. Sometimes we use
the convention $\A_{i,j}:=\C\one$ if $i>j$.
The net $\{\A(I)\}$ is called {\it additive} if $\A(I)$ is
generated by $\{\A_i\,|\,i\in I\}$.

The local observable algebras $\{\A(I)\}$ of the Hopf spin model
defined in subsection 2.1 provide an example of a local additive
split net (see Eqns (2.2--5).) What is not so obvious that this
net satisfies algebraic Haag duality. This follows from the
$\eta$-property described below.

If $\Lambda$ is finite we can choose a system
$\{E_{\alpha}^{ab}\}$ of matrix units for $\A(\Lambda)$ and
define the map
$$\eta_{\Lambda}(A):=\sum_{\alpha}{1\over n_{\alpha}}
\sum_{a,b=1}^{n_{\alpha}}\
E_{\alpha}^{ab}AE_{\alpha}^{ba}\eqno(2.9)$$
which can be seen to be a conditional expectation
$\eta_{\Lambda}\colon\A\to\A(\Lambda)'$. If $tr$ is any trace
state on $\A$ then $tr(B\eta_{\Lambda}(A))=tr(BA)$ for all
$A\in\A$ and $B\in\A(\Lambda)'$. If the net is split then $\A$
is an UHF algebra and there is a unique faithful trace state $tr$ on
$\A$. Hence $\eta_{\Lambda}$ is the orthogonal projection onto
$\A(\Lambda)'$ with respect to the Hilbert--Schmidt scalar
product $<A|B>=trA^*B$. In order to prove Haag duality we need a
kind of orthogonality between the "hyperplanes" $\A(I)$ and
$\A(J)'$ if none of the intervals $I$ and $J$ contains the other.
More precisely we need the

$$\hbox{\sl $\eta$-property }::\qquad\quad
\eqalign{&\eta_i(\A_{i+1,j})\ =\ \A_{i+2,j}\qquad i<j\cr
&\eta_i(\A_{j,i-1})\ =\ \A_{j,i-2}\qquad i>j\cr}$$
where $\eta_i\colon\A\to\A_i'$ is
the conditional expectation defined in (2.9).

\medskip
{\bf Proposition 2.1. } Let $\{\A(I)\}$ be a local net satisfying
the $\eta$-property. Then the net satisfies Haag duality and wedge
duality, i.e.
$$\eqalign{\A(I')'\ &=\ \A(I)\cr
           \A(W')'\ &=\ \A(W)\cr}$$
for all intervals $I\in\I$ and for all wedge regions
$W=\{i,i+1,\dots\}$ or $W=\{\dots,i-1,i\}$.
\smallskip
{\it Proof}: For $\Lambda$ being a wedge $W$ or an
interval $I\in\I$ we can define an $\eta_{\Lambda'}$ as follows:
$$\eqalign{
&\eta_{-\infty,i}(A):=\lim_{j\to-\infty}\ \eta_i\circ\eta_{i-1}
\circ\dots\circ\eta_j(A)\cr
&\eta_{i,\infty}(A):=\lim_{j\to\infty}\ \eta_i\circ\eta_{i+1}
\circ\dots\circ\eta_j(A)\cr
&\eta_{I'}(A):=\eta_{-\infty,i}\circ\eta_{j,\infty}(A)\qquad\qquad
{\rm  if }\ I=\{i+2,\dots,j-2\}\cr
}\eqno(2.10)$$
These infinite products of $\eta_k$-s exists on strictly local
operators $A\in\A_0:=\cup_I\A(I)$ because the sequences under
$\lim_j$ become eventually constants. Now the $\eta$-property
implies that
$\eta_{\Lambda'}(\A_0)\subset\A(\Lambda)$. Since each $\eta_k$
is positive and $\eta_k(\one)=\one$, the same hold for their
limits $\eta_{\Lambda'}$. Hence $\eta_{\Lambda'}$ is continuous
and can be extended to $\A$. The extension also satisfies (2.10)
by an $\epsilon/3$-argument.

If $B\in\A(\Lambda')'$ then for $k\in\Lambda'$ we have
$B\in\A_k'$ therefore $\eta_k(B)=B$. Since $\eta_{\Lambda'}$ is a
product of $\eta_k$-s with $k\in\Lambda'$, we find that
$\eta_{\Lambda'}(B)=B$. This proves
$\A(\Lambda')'\subset\A(\Lambda)$.

If $B\in\A(\Lambda)$ then by locality $B\in\A(\Lambda')'$. This
proves $\A(\Lambda)\subset\A(\Lambda')'$. \hfill {\it Q.e.d.}
\smallskip
In order to apply this result to the Hopf spin model we need to
show that the $\eta$-property holds true in this case. The crossed
product
structure of the local algebras $\A_{i,j}$ imply that every
$A\in\A_{i,j}$ is a linear combination of monomials
$$A=A_iA_{i+1}\dots A_j\qquad {\rm where}\
A_k\in\A_k\,.\eqno(2.11)$$
In this situation the $\eta$-property is equivalent to
$$\eta_i(\A_{i\pm 1})\ =\ \C\one\ .\eqno(2.12)$$
Let us prove (2.12) for $i=$even. (For odd $i$-s the proof is
quite analogous.) Choose $C^*$-matrix units $e_r^{ab}$ of
the algebra $H$. Then one can show that
the coproduct of the integral (see Appendix A) $z=e_0$ takes the form
$$\Delta(z)=\sum_r{1\over n_r}\sum_{a,b}\ e_r^{ab}\otimes e_{\bar
r}^{ab}\eqno(2.13)$$
from which one recognizes that $\eta_i$ evaluated on $\A_{i\pm
1}$ is nothing but
the adjoint action of the integral $z$ on the dual Hopf algebra
$\hat H$. Consider the case of $\A_{i+1}$:
$$\eqalign{
\eta_i(A_{i+1}(\varphi))&=\sum_r{1\over n_r}\sum_{a,b}
\ A_i(e_r^{ab})A_{i+1}(\varphi)A_i(e_r^{ba})\cr
&=A_i(S(z_{(1)}))A_{i+1}(\varphi)A_i(z_{(2)})=A_i(S(z_{(1)})z_{(2)})
A(\varphi_{(2)})\langle\varphi_{(1)}|z_{(3)}\rangle\cr
&=A_{i+1}(\varphi\leftarrow z)=\one\langle\varphi|z\rangle\cr}$$
The case of $\A_{i-1}$ can be handled similarly. This concludes
the proof of the $\eta$-property for the Hopf spin model.

Summarizing: The local net $\{\A(I)\}$ of the Hopf spin model is
an additive split net satisfying Haag duality and
wedge duality. Furthermore the global observable algebra $\A$
is simple, because the split property implies that $\A$
is an UHF algebra and every UHF algebra is simple [Mu].
\bigskip

{\bf 3. Amplimorphisms and comodule algebra actions}
\medskip
{\sl 3.1. The categories $\Amp\A$ and $\Rep\A$}
\medskip
In this subsection $\{\A(I)\}$ denotes a split net of
finite dimensional $C^*$-algebras which satisfies algebraic Haag
duality. Furthermore we assume that the the net is {\it
translation covariant}. That is the net is equipped with a
*-automorphism $\alpha\in\Aut\A$ such that
$$\alpha(\A(I))=\A(I+2)\qquad
I\in\I\,.\eqno(3.1)$$
At first we recall some notions introduced in [SzV].
An {\it amplimorphism} of $\A$ is an injective $C^*$-map
$$\mu\colon\A\to\A\otimes\hbox{End}V\eqno(3.2)$$
where $V$ is some finite dimensional Hilbert space.
If $\mu(\one)=\one\otimes 1_V$  then $\mu$ is called {\it unital}.
Here we will restrict ourselves to unital amplimorphisms since the
localized amplimorphisms in a split net are all
equivalent to unital ones (see Thm. 4.13 in [SzV]). An
amplimorphism $\mu$ is called {\it localized} within $I\in\I$ if
$$\mu(A)=A\otimes 1_V\qquad A\in\A(I^c)\eqno(3.3)$$
where $I^c:=\Z\setminus I$. The space of {\it intertwiners} from
$\nu\colon\A\to\A\otimes\E W$ to $\mu\colon\A\to\A\otimes\E V$
is
$$(\mu|\nu):=\{\,T\in\A\otimes\hbox{Hom}(W,V)\,|\,\mu(A)T=T\nu(A),
\ A\in\A\,\}\eqno(3.4)$$
$\mu$ and $\nu$ are called {\it equivalent}, $\mu\sim\nu$, if
there
exists an isomorphism $U\in(\mu|\nu)$, that is an intertwiner $U$
satisfying $U^*U=\one\otimes 1_W$ and $UU^*=\one\otimes 1_V$.
Let $\mu$ be localized within $I$. Then $\mu$ is called {\it
transportable}
if for all integer $a$ there exists a $\nu$ localized within
$I+2a$ and such that $\nu\sim\mu$. $\mu$ is called $\alpha$
{\it -covariant} if
$(\alpha^a\otimes\id_V)\circ\mu\circ\alpha^{-a}\ \sim\ \mu$ for
all $a\in \Z$.

Let $\Amp\A$ denote the category with objects the localized
unital amplimorphisms $\mu$ and with arrows from
$\nu$ to $\mu$ the intertwiners $T\in(\mu|\nu)$.
This category has the following {\it monoidal product} :
$$\eqalign{
\mu,\ \nu\ \mapsto\ \mu\times\nu&:=(\mu\otimes\id_{End\,
W})\circ\nu\ \colon\ \A\to\A\otimes \E V\otimes\E W\cr
T_1\in(\mu_1|\nu_1), T_2\in(\mu_2|\nu_2)&\mapsto
T_1\times T_2:=(T_1\otimes
1_{V_2})(\nu_1\otimes\id_{\Hom(W_2,V_2)})(T_2)\cr
 &\in\ (\mu_1\times\mu_2|\nu_1\times\nu_2)\cr
}\eqno(3.5)$$
with the monoidal unit being the trivial amplimorphism
$\id_{\A}$. The monoidal product $\times$ is a bifunctor
therefore we have $(T_1\times T_2)(S_1\times
S_2)=T_1S_1\times T_2S_2$, for all intertwiners for which the
products are defined, and $1_{\mu}\times 1_{\nu}=1_{\mu\times\nu}$
where $1_{\mu}:=\one\otimes 1_V$ is the unit arrow at the object
$\mu:\A\to\A\otimes\E V$.

$\Amp\A$ contains {\it direct sums} $\mu\oplus\nu$ of any two
objects: $\mu\oplus\nu(A):=\mu(A)\oplus\nu(A)$ defines a direct
sum for any orthogonal direct sum $V\oplus W$.

$\Amp\A$ has {\it subobjects}: If $P\in(\mu|\mu)$ is a Hermitean
projection then there exists an object $\nu$ and an injection
$S\in(\mu|\nu)$ such that $SS^*=P$ and $S^*S=1_{\nu}$.
The existence of subobjects is a trivial statement in the category
of all, possibly non-unital, amplimorphisms because $S$ can be
chosen to be $P$ in that case. In the category $\Amp\A$ this is a
non-trivial theorem which can be proven [SzV] provided the net is
split. An amplimorphism $\mu$ is called {\it irreducible} if the
only (non-zero) subobject of $\mu$ is $\mu$. Equivalently, $\mu$
is irreducible if $(\mu|\mu)=\C1_{\mu}$. Since the selfintertwiner
space $(\mu|\mu)$ of any localized amplimorphism is finite
dimensional (use Haag duality to show that any $T\in(\mu|\mu)$
belongs to $\A(I)\otimes\E V$ where $I$ is the interval where
$\mu$ is localized), the category $\Amp\A$ is {\it fully
reducible}. That is any object is a finite direct sum of
irreducible objects.

The full subcategory $\Amp^{tr}\A$ of transportable amplimorphisms
is a {\it braided category}. The braiding structure is
provided by the {\it statistics operators}
$$\epsilon(\mu,\nu)\ \in\ (\nu\times\mu|\mu\times\nu)\eqno(3.6)$$
defined by
$$\epsilon(\mu,\nu):=(U^*\o\one)(\onne\o
P)(\mu\o\id)(U)\eqno(3.7)$$
where $U$ is any isomorphism from $\nu$ to some $\tilde\nu$ such
that the localization region of $\tilde\nu$ lies to the left from
that of $\mu$. The statistics operator satisfies
$$\eqalignno{
&\hbox{naturality:}\quad\epsilon(\mu_1,\mu_2)\ (T_1\times T_2)\ =\
(T_2\times T_1)\ \epsilon(\nu_1,\nu_2)&(A.8a)\cr
&\hbox{pentagons:}\quad\eqalign{
 &\epsilon(\lambda\times\mu,\nu)=
  (\epsilon(\lambda,\nu)\times 1_{\mu})(1_{\lambda}\times
  \epsilon(\mu,\nu))\cr
 &\epsilon(\lambda,\mu\times\nu)=
  (1_{\mu}\times\epsilon(\lambda,\nu))(\epsilon(\lambda,\mu)\times
  1_{\nu})\cr}&(3.8b)\cr
}$$
The relevance of the category $\Amp\A$ to the representation
theory of the observable algebra $\A$ can be summarized in the
following theorem taken over from [SzV].
\smallskip
{\bf Theorem 3.1. } Let $\pi_0$ be a faithful irreducible
representation of $\A$ that satisfies Haag
duality:
$$\pi_0(\A(I'))'=\pi_0(\A(I))\qquad I\in\I\,.\eqno(3.9)$$
Let $\Rep\A$ be the category of
representations $\pi$ of $\OA$ that satisfy the following
selection criterion (analogue of the DHR-criterion):
$$\exists I\in\I,\
n\in\N\,:\qquad\pi\vert_{\A(I')}\simeq
n\cdot\pi_0\vert_{\A(I')}\eqno(3.10)$$
where $\simeq$ denotes unitary equivalence. Then $\Rep\A$ is
isomorphic to $\Amp\A$.
If we add the condition that $\pi_0$ is $\alpha$-covariant
and denote by $\Rep^{\alpha}\A$ the full subcategory in $\Rep\A$
of $\alpha$-covariant representations then $\Rep^{\alpha}\A$ is
isomorphic to the category $\Amp^{\alpha}\A$ of $\alpha$-covariant
amplimorphisms.

In general $\Amp^{\alpha}\A\subset\Amp^{tr}\A\subset\Amp\A$.
In the Hopf spin model we shall see that $\Amp^{\alpha}\A=\Amp\A$
(Section 5) and that $\Amp\A$ is equivalent to $\Rep\D(H)$
(subsection 3.5).

\medskip
{\sl 3.2. Comodule algebra actions}
\smallskip
Let $\{\mu_r\}$ be a list of amplimorphisms in $\Amp\A$
containing exactly one from each equivalence class of irreducible
objects. Then an object $\rho$ is called {\it universal} if it is
equivalent to $\oplus_r\mu_r$. Define the $C^*$-algebra $\G$ by
$$\G:=\oplus_r\ \E V_r\eqno(3.11)$$
then every universal object is a unital $C^*$-map $\rho\colon
\OA\to \OA\otimes\G$. The identity morphism $\idA$, as a subobject
of $\rho$, determines
a distinguished 1-dimensional block $\End V_0\cong\C$ of $\G$ and
also a $^*$-algebra map $\varepsilon\colon\G\to\C$.

Universality of $\rho$ implies that the monoidal product
$\rho\times\rho$ is quasiequivalent to $\rho$. The question is
whether there exists an appropriate choice of $\rho$
such that $\rho\times\rho=(\idA\o\Delta)\circ\rho$ for some
coproduct $\Delta\colon\G\to\G\o\G$. If $\rho$ can be chosen in
such a way --- which is probably the characteristic feature of
Hopf algebra symmetry --- then we arrive to the very useful notion
of a comodule algebra action.
\smallskip

{\sl Definition 3.2.}: Let $\G$ be a $C^*$-bialgebra with
coproduct
$\Delta$ and counit $\varepsilon$. A {\it comodule algebra action}
(or shortly a {\it coaction})
on $\A$ is an amplimorphism $\rho\colon\A\to\A\o\G$ that is also
a comodule action on $\A$ with respect to the coalgebra
$(\G,\Delta,\varepsilon)$. In other words: $\rho$ is a linear map
satisfying the axioms:
$$\eqalignno{
\rho(A)\rho(B)&=\rho(AB)&(3.12a)\cr
\rho(\onne)&=\onne\o\one&(3.12b)\cr
\rho(A^*)&=\rho(A)^*&(3.12c)\cr
\exists I\in\I\ :\ \rho(A)&=A\o\one\quad A\in\A(I^c)&(3.12d)\cr
\rho\times\rho\equiv(\rho\otimes\id)\circ\rho\
&=\ (\id\o\Delta)\circ\rho&(3.12e)\cr
(\idA\o\varepsilon)\circ\rho&=\idA&(3.12f)\cr}
$$
$\rho$ is said to be {\it universal} if it is --- as an
amplimorphism --- a universal object of $\Amp\A$.

Examples of comodule algebra actions for the Hopf spin
chain will be given in subsection 3.3. Later, in Sect.5, we will
show that those comodule algebra actions are actually universal.

Every comodule algebra action $\rho\colon\A\to\A\o\G$ determines
an action of the dual $\hat G$ on $\A$, also denoted by $\rho$, as
follows

$$\eqalign{
&\rho_{\xi}\colon\A\to\A\qquad\xi\in\hat\G\cr
&\rho_{\xi}(A):=(\idA\o \xi)(\rho(A))\cr}\eqno(3.13)$$
The axioms for a localized action of the Hopf
algebra $\hat\G$ on the $C^*$-algebra $\A$, that is
$$\eqalignno{
\rho_{\xi}(AB)&=\rho_{\xi_{(1)}}(A)\rho_{\xi_{(2)}}(B)&(3.14a)\cr
\rho_{\xi}(\onne)&=\varepsilon(\xi)\onne&(3.14b)\cr
\rho_{\xi}(A)^*&=\rho_{\xi_*}(A^*)&(3.14c)\cr
\exists I\in\I\ :\ \rho_{\xi}(A)&=A\varepsilon(\xi)\quad
A\in\A(I^c)&(3.14d)\cr
\rho_{\xi}\circ\rho_{\eta}&=\rho_{\xi\eta}&(3.14e)\cr
\rho_{\one}&=\idA&(3.14f)\cr
 }$$
are equivalent to the condition that
$$A\mapsto\rho(A)=\sum_s\ \rho_{\eta_s}(A)\o Y^s\ \in\ \A\o\G$$
is a comodule algebra action, where $\{\eta_s\}$ and $\{Y^s\}$
denote a pair of dual bases of $\hat\G$ and $\G$, respectively.
In (3.14c) we used the notation $\xi\mapsto\xi_*$ for the
antilinear involutive algebra automorphism defined by
$\langle\xi_*|a\rangle=\overline{\langle\xi|a^*\rangle}$. It is
related to the antipode by $\xi_*=S(\xi^*)$.

One can also check that $\rho_{\xi}$ for $\xi=D_r^{kl}$, the
representation matrix of the unitary irrep $r$, determines an
ordinary matrix amplimorphism $\rho_r\colon\A\to\A\o M_{n_r}$.
Whether such a $\rho_r$ is irreducible is not clear for the moment
so we will call it a {\it component} of $\rho$.

For a $\G$-comodule algebra action $\rho$ on $\A$ let
$\Ampr\A$ denote the full subcategory of $\Amp\A$ generated
by objects that are direct sums of irreducibles occuring in $\rho$
as a subobject.

If the bialgebra $\G$ also possesses an antipode $S$ such that
$(\G,\Delta,\varepsilon,S)$ is a $C^*$-Hopf algebra then the
category $\Ampr\A$ becomes equipped with a rigidity
structure: Identifying $\G$ with $\oplus_r\End V_r$ and choosing
evaluation and coevaluation maps $\ev\colon V\o V\to\C$,
$\coev\colon\C\to V\o V$, in the category of vector spaces, we
obtain a transposition map
$$\eqalign{^T\colon\End V&\to\End V\,,\cr
  x\in\End V\ &\mapsto\ x^T:=(1_V\o\ev)(1_V\o x\o 1_V)(\coev\o
  1_V)\cr}$$
such that
$$\bar\rho:=(\idA\o\ ^T\circ S)\circ\rho$$
defines a conjugation functor in $\Ampr\A$ with
evaluation and coevaluation $\Ev=\onne\o\ev$ and
$\Coev=\onne\o\coev$, respectively.

The extremely simple form of the intertwiners $\Ev$ and $\Coev$,
and also of the simple form of the basic intertwiners
$T_{pg}^{r\alpha}=\onne\o t_{pq}^{r\alpha}$ from the
component $\rho_r$ of $\rho$ to the product $\rho_p\times\rho_q$
--- as it is suggested by (3.12e) --- is in fact a general
phenomenon of all amplimorphisms localized within an interval of
length 2, provided the net $\A$ satisfies Haag duality and the
intersection property. The latter one is defined as follows.
\smallskip
{\it Definition 3.3.}: The net $\{\OA(I)\}$ is said to satisfy
the {\it intersection property}  if
$$I,J\in\I,\ I\cap J=\emptyset\ \Rightarrow\
\OA(I)\cap\OA(J)=\C\one\ .\eqno(3.15)$$

Comodule algebra actions with a fixed augmented
$C^*$-algebra $(\G,\varepsilon)$ but with varying coproduct will
be denoted as a pair $(\rho,\Delta)$. In order to compare such
comodule algebra actions one can introduce equivalences of three
different kinds.

{\it Definition 3.4.}: Let $(\rho,\Delta)$ and $(\rho',\Delta')$
be comodule algebra actions of $(\G,\varepsilon)$ on $\A$. Then a
pair $(U,u)$ of unitaries $U\in\A\o\G$ and $u\in\G\o\G$ is called
a {\it cocycle equivalence} from $(\rho,\Delta)$ to
$(\rho',\Delta')$ if
$$\eqalignno{
U\rho(A)&=\rho'(A)U\qquad A\in\A&(3.16a)\cr
u\Delta(X)&=\Delta'(X)u\qquad X\in\G&(3.16b)\cr
U\times U&=(\onne\o u)\cdot(\idA\o\Delta)(U)&(3.16c)\cr}
$$
$(U,u)$ is called a {\it coboundary equivalence} if in addition to
(a--c)
$$u=(x^{-1}\o x^{-1})\Delta(x)\eqno(3.16d)$$
holds for some unitary $x\in\G$. If $u=\one\o\one$, $\rho$ and
$\rho'$ are called {\it equivalent}.

Notice that as a consequence of (3.16c) the unitary $u$ satisfies
the cocycle condition
$$(\one\otimes
u)(\id\otimes\Delta)(u)=(u\otimes\one)(\Delta\otimes\id)(u)\eqno(3.17)$$
which is precisely the condition for a twisting to
preserve quasitriangularity of an $R$-matrix: $R\mapsto
R'=u^{op}Ru^{-1}$.

{\bf Lemma 3.5. } {\sl Let $\{\A(I)\}$ be a net satisfying
algebraic Haag duality and intersection property. Let furthermore
$\{\rho_r\}$ be a finite family of irreducible amplimorphisms
$\rho_r\colon\A\to\A\o\End V_r$,
closed under the monoidal product, and such that each one of them
is localized within the same interval $I$ of length $|I|=2$.
Let $\rho=\oplus_r\rho_r$ be localized within $I$, too. Then there
exists one and only one coproduct $\Delta$ on $\G=\oplus_r\End
V_r$ such that the pair $(\rho,\Delta)$ is a comodule algebra
action of $\G$ on $\A$.}

{\it Proof}: Haag duality and intersection property imply that all
intertwiners in $T\in(\rho_p\times\rho_q|\rho_r)$ are scalars,
i.e. have the form $T=\onne\o t$ with some $t\colon V_r\to V_p\o
V_q$. Choose an orthonormal basis $\{t_{pq}^{r\alpha}\}$ in
$\Hom(V_r,V_p\o V_q)$ and define
$$\Delta(X):=\sum_{pqr\alpha}\
t_{pq}^{r\alpha}\,X\,t_{pq}^{r\alpha\,*}\,,\qquad X\in\G$$
It is easy to verify that this map defines a coproduct on $\G$ and
furthermore that the intertwiners $T_{pq}^{r\alpha}:=\onne\o
t_{pq}^{r\alpha}$ are complete and
orthonormal in $(\rho_p\times\rho_q|\rho_r)$. Therefore
$\rho\times\rho=(\idA\o\Delta)\circ\rho$ follows.

Now assume that there exists an other coproduct $\Delta'$ on $\G$
such that $(\rho,\Delta')$ is also a comodule algebra action.
Then the "fusion coefficients" $N_{pq}^r$ of $\Delta'$ must be the
same as those of $\Delta$ since both of them are determined by the
composition rules of the irreducible objects $\rho_r$. Therefore
there exists a twisting $u\in\G\o\G$ such that
$\Delta'=\Ad_u\circ\Delta$. Now multiplying the identity
$$(\idA\o\Delta')\circ\rho=(\idA\o\Delta)\circ\rho$$
by $ut_{pq}^{s\beta}$ from the right and by $t_{pq}^{r\alpha\,*}$
from the left we obtain
$$\onne\o t_{pq}^{r\alpha\,*}ut_{pq}^{s\beta}\ \in\ (\rho|\rho)$$
a selfintertwiner. Since $\rho$ contains every irreducible only
once, this selfintertwiner is a central element of $\G$. That is
there exist complex numbers $c_{pq}^{r\alpha\beta}$ such that
$$\eqalign{
  t_{pq}^{r\alpha\,*}&ut_{pq}^{s\beta}=\delta^{rs}
  c_{pq}^{r\alpha\beta}\cdot e_r,\qquad (e_r=\id_{V_r})\cr
  &u=\sum_{pqr}\sum_{\alpha\beta}\
t_{pq}^{r\alpha}\,c_{pq}^{r\alpha\beta}\,t_{pq}^{r\beta\,*}\cr
}$$
Now it is easy to check that this $u$ commutes with all
$\Delta(X)$, hence $\Delta'=\Delta$.\hfill{\it Q.e.d.}
\smallskip
{\bf Proposition 3.6. } {\sl Let $\{\A(I)\}$ be as in the previous
Lemma. Let $(\rho_0,\Delta_0)$ be a comodule algebra action of the
augmented algebra $(\G,\varepsilon)$ that is localized within an
interval of length 2. Then universal comodule algebra
actions of ${\bf Amp}_{\rho_0}$ are unique up to cocycle
equivalence.}

{\it Proof}: Let $(\rho,\Delta)$ be a universal comodule algebra
action of ${\bf Amp}_{\rho_0}$. Then $\exists
U\in\Iso(\rho_0|\rho)$. We have then two isometric intertiners
$$\eqalign{
(\idA\o\Delta)(U)&\colon\rho\times\rho\rightarrow(\idA\o\Delta)
                  \circ\rho_0\cr
U\times U&\colon\rho\times\rho\rightarrow\rho_0\times\rho_0\cr
}$$
Since both $(\idA\o\Delta)\circ\rho_0$ and $\rho_0\times\rho_0$ is
localized within the same interval of length 2, the intertwiner
$(U\times U)\cdot (\idA\o\Delta)(U^*)$ between them must be a
scalar: $\onne\o u$. We obtain
$$\eqalign{
U\times U&=(\onne \o u)\cdot (\idA\o\Delta)(U)\cr
\rho_0\times\rho_0&=(\idA\o\Delta')\circ\rho_0\cr
\hbox{where}\ \ \Delta'&=\Ad_u\circ\Delta\cr
}$$
This proves that $(U,u)$ is a twisted equivalence from
$(\rho,\Delta)$ to $(\rho_0,\Delta')$. By Lemma 3.5 $\Delta'$ is
equal to $\Delta_0$, therefore every universal comodule algebra
action is cocycle equivalent to the same $(\rho_0,\Delta_0)$.
\hfill{\it Q.e.d.}

\medskip
{\sl 3.3. The special comodule actions and their charge
transporters}
\smallskip
Let $\G$ denote the Drinfeld Double $\D(H)$ (See Appendix B).
The formulae given below define amplimorphisms
$\rho_I\colon A\to\A\o
\G$ that are localized on an interval $I$ of length 2:
$$\eqalign{
\reven(A_{2i}(a)A_{2i+1}(\varphi))&:=A_{2i}(a_{(1)})
A_{2i+1}(\varphi_{(2)})\ \o\ \D(a_{(2)})\D(\varphi_{(1)})\cr
\rodd(A_{2i-1}(\varphi)A_{2i}(a))&:=A_{2i-1}(\varphi_{(1)})
A_{2i}(a_{(2)})\ \o\ \D(\varphi_{(2)})\D(a_{(1)})\cr}
\eqno(3.18)$$
The proof of that these expressions really determine
amplimorphisms is straightforward
and will be omitted. Likewise we left to the reader to check that
(3.18) in fact define comodule algebra actions, that is
$$\left.\eqalign{
\reven\ \times\ \reven\ &=\ (\idA\otimes\Del)\circ\reven\cr
\rodd\ \times\ \rodd\ &=\ (\idA\otimes\Delop)\circ\rodd\cr
(\idA\otimes\varepsilon_{\D})\circ\rho_{i,i+1}\ &=\ \idA\cr
}\right\}\eqno(3.19)$$
Hence $\reven$ is a comodule action with respect to the coalgebra
$\D(H)=$\break$(\G,\Del,\varepsilon_{\D})$ and $\rodd$ is one with
respect
to $\D(\hat H)=(\G,\Delop,\varepsilon_{\D})$. The formulae (3.18)
are manifestly translation covariant, so we have
$$(\alpha\o\idG)\circ\rho_{i,i+1}\circ\alpha^{-1}
\ =\ \rho_{i+2,i+3}\eqno(3.20)$$
Of course, one expects that amplimorphisms related by translations
(3.20) are actually equivalent, therefore there exists appropriate
charge transporters between them. It is not
clear, however, whether the amplimorphisms $\reven$ and $\rodd$
create independent sectors or not. Let us define the charge
transporter $T_i$ as follows:
$$T_i:=\left\{\eqalign{
&A_i(b_A)\otimes \D(\beta^A)\qquad i=\hbox{even}\cr
&A_i(\beta^A)\otimes \D(b_A)\qquad i=\hbox{odd}\cr
}\right.\eqno(3.21)$$
Then we have

{\bf Proposition 3.7. }{\sl The charge transporters $T_i$ are
unitary intertwiners from $\rho_{i,i+1}$ to $\rho_{i-1,i}$, i.e.
$$T_i\rho_{i,i+1}(A)\ =\ \rho_{i-1,i}(A)T_i\,,\qquad
A\in\A\eqno(3.22)$$
and satisfy the cocycle condition
$$\eqalign{
T_i\times T_i\ &\equiv\
(T_i\otimes\one)\cdot(\rho_{i,i+1}\otimes\id)(T_i)\ =\cr
&=\left\{\eqalign{
&(\one\otimes R)\cdot(\id\o\Del)(T_i)\quad i=\hbox{even}\cr
&(\one\o R^{op})\cdot(\id\o\Delop)(T_i)\quad i=\hbox{odd}\cr}
\right.\cr}\eqno(3.23)$$
}
{\it Proof}: By inspection.
\medskip
In view of Definition 3.4 the above proposition claims that the
pair $(T_i,R^{(op)})$ determines a cocycle equivalence between
comodule actions of the form \break
 $(\rho_{i,i+1},\Delta_D^{(op)})$.
More precisely, we have the infinite sequence of cocycle
equivalences
$$\dots(\reven,\Del)
\stackrel{(T_{2i+1},R^{op})}{\longleftarrow}
(\rho_{2i+1,2i+2},\Delop)
\stackrel{(T_{2i+2},R)}{\longleftarrow}
(\rho_{2i+2,2i+3},\Del)\dots
\eqno(3.24)$$
Composing these two arrows we obtain a coboundary
equivalence\hfill\break
$(T_{2i+1}T_{2i+2}, R^{op}R)$ because $R^{op}R=(s\o
s)\Del(s^{-1})$ according to (A.7) where $s$ is given by (B.6).
Likewise $(T_{2i}T_{2i+1},RR^{op})$ yields a coboundary
equivalence. Therefore introducing
 $$U_{i,i+1}:=(\one\o s^{-1})T_iT_{i+1}\quad\in\ (\rho_{i-1,i}|
 \rho_{i+1,i+2})\eqno(3.25)$$
we obtain a charge transporter localized within $\{i,i+1\}$ that
satisfies the cocycle condition
$$\eqalign{
U_{2i+1,2i+2}\ \times\ U_{2i+1,2i+2}\ &=\ (\idA\o
\Del)(U_{2i+1,2i+2})\cr
U_{2i,2i+1}\ \times\ U_{2i,2i+1}\ &=\ (\idA\o
\Delop)(U_{2i,2i+1})\cr
}\eqno(3.26)$$
The existence of such charge transporters means --- by definition
--- that the comodule actions $\rho_{i,i+1}$ are
$\alpha$-covariant (and not only transportable).

If we want to see the components of the
amplimorphism $\rho_{i,i+1}$ we can proceed as follows. Choose a
system $\{E_r^{kl}\}$ of $C^*$-matrix units for $\G$ and dual
basis $\{D_r^{kl}\}$ for $\hat \G$, i.e.
 $$\eqalignno{
 E_p^{ij}E_q^{kl}\ &=\ \delta_{pq}\delta^{jk}E_p^{il}
 \qquad (E_p^{kl})^*\ =\ E_p^{lk}&(3.27a)\cr
 \langle D_p^{ij},E_q^{kl}\rangle\ &\equiv\ D_p^{ij}(E_q^{kl})\ =\
 \delta_{pq}\delta^{ik}\delta^{jl}&(3.27b)\cr
 \Delta(D_r^{kl})\ &=\ \sum_m\ D_p^{km}\o D_p^{ml}&(3.27c)\cr}$$
Then introducing the notation
$$\rho_r^{kl}:=\rho_{D_r^{kl}}\eqno(3.28)$$
one can verify that
$$\eqalign{
\rho_r^{kl}(AB)=\rho_r^{km}(A)\rho_r^{ml}(B)\cr
\rho_r^{kl}(A)^*=\rho_r^{lk}(A^*)\cr}\eqno(3.29)$$
that is $\rho_r\colon\A\to\A\o M_{n_r}$ is a $^*$-algebra map.
One expects, of course, that $\rho_r$ is irreducible and
that the intertwiners in $(\rho_p\times\rho_q|\rho_r)$ are in
one-to-one correspondence with the intertwiners in $(D_p\times
D_q|D_r)$. This is, however, not so trivial and we will return to
it in subsection 3.5.
\medskip
{\sl 3.4. Outerness of $\rho$}
\smallskip
Let $\rho$ denote one of the comodule actions
defined in (3.18). Then the corresponding action $\rho_{\xi}=
(\idA\o\xi)\circ\rho$
of $\hat\G$ is faithful in the following sense:
$$\rho_{\xi}(A)=0\quad\forall A\in\A\quad\Rightarrow\quad\xi=0
\eqno(3.30)$$

We prove the statement for the case $\rho=\reven$. Assume
$$\rho_{\xi}(A_{2i}(a)A_{2i+1}(\varphi)=
A_{2i}(a_{(1)})A_{2i+1}(\varphi_{(2)})\ \o\ \xi(\D(a_{(2)})
\D(\varphi_{(1)}))\ =\ 0$$
for some $\xi\in\hat\G$ and for all $a\in H,\ \varphi\in\hat H$.
Multiplying this equation from the left by
$A_{2i+1}(\varphi_{(3)})$$A_{2i}(a_{(0)})$ we obtain
$$\one\ \o\ \xi(\D(a)\D(\varphi))\ =\ 0\ $$
which immediately implies that $\xi=0$.

Using faithfulness of the action $\rho_{\xi}$ we can completely
determine the selfintertwiner space $(\rho|\rho)$ as follows.

Notice at first that for
computing intertwiners it is not enough to know $\rho$ as a map to
the abstract $C^*$-algebra $\A\o\G$ but we have to specify $\G$ as
a $^*$-subalgebra of some $\End V$. Our convention is that $\G$ is
embedded into $\End V$ as its defining representation, i.e. as the
direct sum of its irreducible representations each of them with
multiplicity 1. This remark completes the definition of the
special comodule algebra actions of the doubles $\G=\D(H)$
or $\G=\D(\hat H)$, respectively, given in (3.18).

With this definition a selfintertwiner $T$ of a $\rho$ of (3.18)
is neccesarily a scalar, i.e. has the form $T=\onne\o t$
where $t\in\End V$, because $\rho$ is localized on a 2-point
interval. The intertwining property
$$(\onne\o t)\cdot\rho(A)\ =\ \rho(A)\cdot(\onne\o t)\qquad
A\in\A$$
can be equivalently written as
$$\rho_{\xi}(A)=0\quad\forall A\in\A$$
for all $\xi\in\hat\G$ such that
$$\xi(X)=(u, (Xt-tX) v)\qquad\hbox{for some $u,v\in V$}\ .$$
Now faithfulness of $\rho$ implies that $Xt-tX=0$ for all
$X\in\G$,
that is $t\in\G'=\G\cap\G'$ by the definition of the embedding
$\G\subset\End V$. This proves that 2-point localized faithful
comodule algebra actions $\rho$ of the $C^*$-Hopf algebra $\G$ on
a Haag dual
net $\A$ with intersection property have selfintertwiner spaces
$$(\rho|\rho)\ =\ \onne\,\o\,\hbox{Center}\,\G\,.\eqno(3.31)$$
In particular the irreducible components of such a $\rho$ are in
one-to-one correspondence with minimal central projectors of $\G$.
That is the components of $\rho$ are precisely its irreducible
components.

The special form (3.31) of the selfintertwiner space implies the
weaker but important property of outerness of $\rho$.

{\it Definition 3.8.}: An amplimorphism $\mu\colon\A\to\A\o M_n$
is
called {\it inner} if there exists a unitary $U\in\A\o M_n$ such
that $U(A\o I_n)U^*=\mu(A)$ for $A\in\A$.
The comodule algebra action $\rho\colon\A\to\A\o\G$ of the
$C^*$-bialgebra $\G$ on the $C^*$-algebra $\A$ is {\it
outer} iff none of the components of $\rho$
is inner except the trivial
amplimorphism $(\idA\o\varepsilon)\circ\rho=\idA$.
The action $\rho\colon\hat \G\o\A\to\A$ is {\it outer} iff the
corresponding coaction is outer.

{\bf Lemma 3.9. } {\sl Let $\A$ be a Haag dual net with
intersection
property and $\rho\colon\A\to\A\o\G$ be a coaction of the Hopf
algebra $\G$ that is localized on an interval $I$ of length 2.
Then faithfulness of the $\hat\G$-action $\rho\colon\G\o\A\to\A$
implies outerness of the (co)action $\rho$.}

{\it Proof}: From (3.31) we it follows that none of the
non-trivial components $\rho_r=(\idA\o D_r)\circ\rho$, $r\neq 0$,
can have intertwiners to the trivial component $\rho_0=\idA$.
Therefore $\rho_r$ is inner only for $r=0$.\hfill{\it
Q.e.d.}

Applying this Lemma to the Hopf spin chain we conclude that the
comodule algebra actions defined in (3.18) are outer.

\vfill\eject

\medskip
{\sl 3.5. The equivalence of $\Ampr\A$ and $\Rep\G$}
\smallskip
We recall that
$\Ampr\A$ denotes the full subcategory of $\Amp\A$
generated by objects $\mu$ of the form $\mu\sim\oplus_s\mu_s$
where each $\mu_s$ is an irreducible subobject of $\rho$.

{\bf Theorem 3.10. } {\sl Let $\A$ be a Haag dual net
satisfying the intersection property. Let $\rho$ be a faithful
comodule
algebra action of the quasitriangular $C^*$-Hopf algebra $\G$ on
$\A$ that is localized on an interval of length 2. Then the category
$\Ampr\A$ and the
category of $^*$-representations $\Rep\G$ of $\G$ are equivalent
as strict monoidal, braided, rigid, $C^*$-categories.}

{\it Proof}: We prove the equivalence $\Ampr\A\sim\Rep\G$ as
equivalence of\break
 monoidal categories. Other features such as
rigidity $\dots$ etc. can be checked rather easily (cf. to the
proof of Thm.4.16. in [SzV]) and therefore are omitted.

We need to construct a monoidal functor
$\Phi\colon\Ampr\A\to\Rep\G$
which is one-to-one on the equivalence classes of objects and is
one-to-one between intertwiner spaces
$(\mu|\nu)\to(\Phi(\mu)|\Phi(\nu))$ for each pair $\mu,\nu$ of
amplimorphisms. Such a functor is readily obtained once we
have established: (I) a one-to-one map $\rho_r\mapsto D_r$ between
a complete family
$\{\rho_r\}$ of irreducible objects of $\Ampr\A$ and a
complete
family $\{D_r\}$ of irreducible objects of $\Rep\G$; and (II)
one-to-one maps $(\rho_p\times\rho_q|\rho_r)\to(D_p\times
D_q|D_r)$ between the basic intertwiner spaces.

Let
$D_r\colon\G\to M_{n_r}$ be fixed irreducible $^*$-representations
of $\G$ and define $\rho_r:=(\idA\o D_r)\circ\rho$. Since each
$\rho_r$ is localized within $\{i,i+1\}$, the intersection
property and Haag duality implies that all intertwiners
$T\in(\rho_p\times\rho_q|\rho_r)$ have the form $T=\one\o t$ with
some $t\in\hbox{Mat}(n_pn_q\times n_r,\C)$. We claim that
$$
\eqalign{
\rho_r&\mapsto D_r\cr
T=\one\o t\in(\rho_p\times\rho_q|\rho_r)&\mapsto t\in(D_p\times
D_q|D_r)\cr}
$$
defines the required functor $\Phi$. Let us show at first that
$t\in(D_p\times D_q|D_r)$:

$$\eqalign{
\rho_p\times\rho_q(A)(\one\o t)&=(\one\o t)\rho_r(A)\cr
(\idA\o D_p\o D_q)\circ(\idA\o\Delta)\circ\rho(A)\cdot(\one\o t)
&=(\one\o t)\cdot(\idA\o D_r)\circ\rho(A)\cr
(\idA\o\xi)\circ\rho(A)\ &=\ 0\qquad A\in\A\cr}$$
for all $\xi\in\hat\G$ which associates to $X\in\G$ one of the
matrix elements of
$$(D_p\times D_q)(X)\cdot t\ -\ t\cdot D_r(X)\,.$$
By faithfulness of $\rho$ all these $\xi$'s must be zero.
Therefore $t\in(D_p\times D_q|D_r)$ as claimed.

In order to see surjectivity of the map $T\mapsto t$ let
$t\in(D_p\times D_q|D_r)$. Then $\one\o t$ is in
$(\rho_p\times\rho_q|\rho_r)$ by the very definition of the
$\rho_p$'s. Hence $T\mapsto t$ is one-to-one.
\hfill{\it Q.e.d.}

\vfill\eject

{\bf 4. Construction of field algebras}
\medskip
Field algebras $\F$ are $C^*$-algebra extensions $\F\supset\A$ of
the observables that satisfy the following physical requirements:

I. $\F$ contains all charge carrying fields: For any localized
amplimorphism $\mu$ there exists $F_{\mu}\in\F\o\End V_{\mu}$ such
that $\F_{\mu}$ implements $\mu$, i.e. $F_{\mu}(A\o\one)=\mu(A)F_{\mu}$
for all $A\in\A$.

II. The inclusion$\A\subset\F$ must be irreducible: $\A'\cap\F=\C\cdot\onne$.
In other words, an operator that commutes with all observables should
be a symmetry and not a field.

III. There exists a conditional expectation ${\cal E}\colon\F\to\A$ of
index finite type. This requirement comes from that we want $\F$ to carry
an action of the quantum symmetry such that $\A$ is the invariant
subalgebra. ${\cal E}$ will then be the quantum group average. The finiteness
of the index is related to our interest in finite dimensional quantum
symmetries.

IV. $\F$ is minimal under conditions (I--III). That is if $\F_1$ satisfies
$\A\subset\F_1\subset\F$ and conditions (I--III) above then $\F_1=\F$.
\medskip

{\sl 4.1. The field algebras $\Fr$ }
\medskip
For a comodule action $(\rho,\Delta)$ we define the
field algebra $\Fr$ as the crossed product
$$\eqalignno{
\Fr=\A\cros\hat \G\ &=\
\hbox{Span}\{AF_{\xi}\,|\,A\in\A,\xi\in\hat \G\,\}&(4.1)\cr
F_{\xi}F_{\eta}&=F_{\xi\eta},&(4.1a)\cr
F_{\xi}^*&=F_{\xi^*},&(4.1b)\cr
F_{\xi}A&=\rho_{\xi_{(1)}}(A)F_{\xi_{(2)}}&(4.1c)\cr}$$

There is an action $\gamma$ of the symmetry algebra $\G$ on $\A$
with respect to which the invariant subalgebra $\F^{\gamma}$ is
precisely the observable algebra:
$$\eqalign{
\gamma_X(AF_{\xi})\ &:=\ AF_{X\rightarrow\xi}\qquad
X\in\G,\ A\in\A,\ \xi\in\hat \G\cr
\F^{\gamma}&\equiv\{F\in\F\,|\,\gamma_{X}(F)=\varepsilon(X)F\,\}\
=\ \A\cr}$$
If $h$ denotes the integral of the Hopf algebra $\G$ then ${\cal E}:=
\gamma_h$ defines a conditional expectation onto $\A$, the ``average''
over $\G$. It can be seen to be of index finite type if $\G$ is finite
dimensional. As a matter of fact let $\{\eta_s\}$ be an orthonormal basis
of $\hat \G$ with respect to the scalar product $\langle\xi|\eta\rangle=
\langle\xi^*\eta,h\rangle$. Then $F_{\eta_s}$ provide us with a quasibasis
[Wa] for the conditional expectation ${\cal E}$, i.e.
$$\sum_s\ F_{\eta_s}\,{\cal E}\left(F_{\eta_s}^*F\right)\ =\ F\qquad
\forall F\in\F\,.$$
For computing the index take into account that $\langle\xi,h\rangle$ is just
the trace of $\xi$ in the left regular representation. Therefore the $\eta_s$
basis consists of appropriately normalized matrix units of $\hat \G$.
This gives $\hbox{Index}{\cal E}=\sum_s F_{\eta_s}F_{\eta_s}^*=
\hbox{dim}\G\cdot\onne$.

An equivalent formulation of the defining relations (4.1.a--c)
can be given by using the "master field"
$$F:=F_{\eta_s}\,\o\,Y^s\ \ \in\ \Fr\o\G\eqno(4.2)$$
as follows:
$$\eqalignno{
F^{01}F^{02}\ &=\ (\id\o\Delta)(F)&(4.2.a)\cr
F^*\ =\ F^{-1}\ &=\ (\id\o S)(F)&(4.2.b)\cr
F(A\o\one)\ &=\ \rho(A)F&(4.2.c)\cr
}$$

Now we turn to the question of how the crossed product depends on
the comodule action.

{\bf Theorem 4.1. } {\sl Two universal outer comodule algebra
actions
$(\rho,\Delta)$ and $(\rho',\Delta')$ of $\G$ on the
observable algebra $\A$ of the Hopf spin model give
rise to isomorphic crossed products $\Fr=\A\cros_{\rho}\G$ and
$\F_{\rho'}=\A\cros_{\rho'}\G$ if and only if
$\rho$ and $\rho'$ are coboundary equivalent.}

Recall that under the "isomorphism of the crossed products $\Fr$
and $\F_{\rho'}$" we mean that there exists a
$C^*$-algebraic isomorphism between them which leaves the
observable algebra $\A$ pointwise invariant.

{\it Proof}: Necessity: Let $(\rho,\Delta)$, $(\rho',\Delta')$ be
universal outer comodule algebra actions such that the
corresponding
crossed products $\Fr$ and $\F_{\rho'}$ are isomorphic. Let
$\theta\colon\F_{\rho'}\to\Fr$ be such an isomorphism and let $F$
and $F'$ denote the master fields of the two field algebras.
Pick up a twisted equivalence
$(U,u)\in\left(\left(\rho',\Delta'\right)|
\left(\rho,\Delta\right)\right)$, which exists by Proposition 3.6,
and construct
$G=F^*U^*(\theta\o\idG)(F')$. One checks easily that $G$ commutes
with all operators of the form $A\o\one$ where $A\in\A$. Thus
$G\in\A'\o\G$. Since outernes of $\rho$ implies that $\A$ has
trivial relative commutant within $\Fr$ (see Proposition 4.2),
it follows that $G=\onne\o x$, with some $x\in\G$. Hence we found
that the two master fields are related by
$$(\theta\o\idG)(F')\ =\ UF(\onne\o x)\,.$$
Now we may compute the operator product
$$\eqalign{
(\theta\o\idG\o\idG)(F'_{01}F'_{02})&=
          (UF)_{01}(UF)_{02}(\onne\o x\o x)=\cr
&=(U\times U)F_{01}F_{02}(\onne\o x \o x)=\cr
&=(\id\o\Delta')(UF)\cdot(\onne\o u(x\o x))=\cr
&=(\theta\o\idG\o\idG)\circ(\id\o\Delta')(F')\cdot
          (\onne\o u\Delta(x^{-1})(x\o x))\cr
          }$$
which immediately implies that $u=(x^{-1}\o x^{-1})\Delta(x)$,
a coboundary.

Sufficency: Let $(U,u)$ be a coboundary equivalence from
$(\rho,\Delta)$ to $(\rho',\Delta')$. Then with $x$ such that
$u=(x^{-1}\o x^{-1})\Delta(x)$ we can define the map
$$\eqalign{
\theta\colon\F_{\rho'}&\to\Fr\cr
\theta\o\idG\colon (A\o\one)F'&\mapsto (A\o\one)UF(\onne\o x)\cr
}$$
Now it is not difficult to verify that $\theta$ is a
$^*$-isomorphism which leaves $\A$ pointwise invariant. What one
needs to do is only to check that $UF(\onne\o x)\in\Fr\o\G$ is a
master field associated to $(\rho',\Delta')$.\hfill{\it Q.e.d.}

\medskip

{\sl 4.2. The irreducibility of the inclusion $\A\subset\F$}
\medskip
If $\F$ had been constructed as a crossed product of $\A$
with the action $\rho$ of a group (instead of $\hat\G$) then we
would conclude that irreducibility of $\A\subset\F$ is equivalent
to the outerness of the action $\rho$: $\rho_g$ is an inner
automorphisms of $\A$ iff $g=1$. In fact this conclusion holds
also for any $C^*$-Hopf algebra action which is outer in the sense
of Definition 3.8.

{\bf Proposition 4.2. } {\sl Let $\A$ be a unital $C^*$-algebra
and let $\F$ be the crossed product of $\A$ with respect to the
coaction $\rho$ of the $C^*$-Hopf algebra $\G$. Then
$\A'\cap\F=\A'\cap\A$ if and only if $\rho$ is outer.
In particular if $\A$ has trivial center then the
inclusion $\A\subset\F$ is irreducible
if and only if the action $\rho$ is outer.}

{\it Proof}: Let $C\in\A'\cap\F$ be a unitary which is not an
observable. The identity
$$[\gamma_x(C),A]=\gamma_{x_{(1)}}(C)\gamma_{x_{(2)}}(A)-
\gamma_{x_{(1)}}(A)\gamma_{x_{(2)}}(C)=\gamma_x([c,A])=0\quad
A\in\A$$
shows that for each $x\in\G\ $ $\gamma_x(C)\in\A'\cap\F$ too.
Since $C$ is not an observable, there exist an $r\neq 0$ such that
$\gamma_{E_{\bar r}}(C)\neq 0$. With this $r$ construct the fields
$$C_r^{ki}:=\sum_{n,j}\ \gamma_{E_{\bar r}^{ni}}(C)F_r
^{kj}(t_{\bar r r}^0)^{nj,.}$$
where $t_{pq}^{r\alpha}\colon D_r\to D_p\times D_q$ denote
orthonormal intertwiners (Clebsh-Gordan maps) for the
representation theory of $\G$. The $F_r$ implements the
amplimorphism $\rho_r$ therefore
$$C_r^{ki}A=\rho_r^{kj}(A)C_r^{ji},\qquad A\in\A\,.$$
On the other hand the $C_r^{ki}$ are observable since
$$\eqalign{
\gamma_x(C_r^{ki})&=\sum_{njl}\ \gamma_{x_{(1)}E_{\bar r}^{ni}}(C)
F_r^{kl}D_r^{lj}(x_{(2)})(t_{\bar rr}^0)^{nj,.}=\cr
&=\sum_{njlm}\ \gamma_{E_{\bar r}^{mi}}(C)F_r^{kl}
  D_{\bar r}^{mn}(x_{(1)})D_r^{lj}(x_{(2)})(t_{\bar
  rr}^0)^{nj,.}=\cr
&=\varepsilon(x)\sum_{lm}\ \gamma_{E_{\bar r}^{mi}}(C)F_r^{kl}
  (t_{\bar rr}^0)^{ml,.}=\cr
&=\varepsilon(x)C_r^{ki}\cr}$$
Therefore $\rho_r$ is inner.

Assume that for some $r\neq 0$ $(\rho_r|\idA)\neq 0$, i.e.
$$\exists C_r\in\A\o M_{n_r}, C_r\neq 0\ \hbox{such that}\
C_r^{ki}A= \rho_r^{kj}(A)C_r^{ji}\ a\in\A$$
Then $S^{ji}:=F_r^{kj*}C_r^{ki}$ satisfies
$S^{ji}A=AS^{ji}$ for all $A\in\A$ hence $S^{ji}\in\A'\cap\F$.
We want to show that the $S^{ji}$ are not observables.
$$\eqalign{
\gamma_X(S^{ji})&=\gamma_X(F_r^{kj*})C_r^{ki}=\left[\gamma_{S(X^*)}
(F_r^{kj})\right]^*C_r^{ki}=\cr
&=\left[F_r^{kl}D_r^{lj}(S(X^*))\right]^*C_r^{ki}=
  D_r^{jl}(S(X))F_r^{kl*}C_r^{ki}=\cr
  &=D_r^{jl}(S(X))S^{li}\qquad\qquad X\in\G\,.\cr}$$
Therefore $\A'\cap\F$ is not contained in $\A$.\hfill{\it Q.e.d.}

As a consequence of this proposition we can conclude that the
field
algebras $\F_{i,i+1}$ associated to the special comodule actions
$\rho_{i,i+1}$ are irreducible.

\medskip
{\sl 4.3. Translation covariance}
\smallskip
The problem of translation covariance of a field algebra extension
consists of showing that the automorphism $\alpha$ of $\A$ extends
to an automorphism $\hat \alpha$ of the crossed
product $\F=\A\cros\G$. Further requirement is that $\hat\alpha$
be geometric, that is $\hat\alpha$ should commute with the
internal symmetry $\gamma$.

{\bf Theorem 4.3. } {\sl Let $\alpha\in\Aut\A$ and let
$(\rho,\Delta)$ be a comodule action. Then there exists an
extension $\hat\alpha\in\Fr$ of $\alpha$ commuting with the action
$\gamma$ of $\G$ if and only if there exists
$U\in\Iso(\rho^{\alpha}|\rho)$ satisfying the cocycle condition
$$U\times U\ =\ (\idA\o\Delta)(U)\ .\eqno(4.3)$$
If this is the case then $\hat\alpha$ is unique up to a central
grouplike unitary $g\in\G$ specified below.}

{\it Proof}: We use the master field notation of (4.2).
If $\hat\alpha$ is an extension then
introducing
$$\eqalignno{
F^{\alpha}&=(\hat\alpha\o\id)(F)&(4.4)\cr
\rho^{\alpha}&=(\alpha\o\idG)\circ\rho\circ\alpha^{-1}&(4.5)\cr
}$$
a little calculation shows that
$$\eqalignno{
F^{\alpha}(A\o\one)&=\rho^{\alpha}(A)F^{\alpha}&(4.6a)\cr
F^{\alpha}_{01}F^{\alpha}_{02}&=(\id\o\Delta)(F^{\alpha})
&(4.6b)\cr
\rho^{\alpha}\times\rho^{\alpha}&=(\idA\o\Delta)(\rho^{\alpha})
&(4.6c)\cr
}$$
Therefore $(\rho^{\alpha},\Delta)$ is also a comodule algebra
action and $F^{\alpha}$ is the associated master field.
Define $U:=F^{\alpha}F^*\in\Fr\o\G$ which is unitary. In order for
this $U$ to belong to $\A\o\G$ we have to assume that $\hat\alpha$
commutes with $\gamma$. As a matter of fact
$$\eqalign{
(\gamma_X\o\id)(F)&=F(\onne\o X)\cr
(\gamma_X\o\id)(F^{\alpha})&=F^{\alpha}(\onne\o X)\cr
(\gamma_X\o\id)(U)&=F^{\alpha}(\onne\o X_{(1)})
                    (\onne\o S(X_{(2)}))F^*=U\varepsilon(X)\cr
}$$
Now it is easy to check that $U\rho(A)=\rho^{\alpha}(A)U$ hence
$U\in\Iso(\rho^{\alpha}|\rho)$. Since $F^{\alpha}=UF$,
$$\eqalign{
 F^{\alpha}_{01}F^{\alpha}_{02}&=(U\times U)F_{01}F_{02}=\cr
     &=(U\times U)(\idA\o\Delta)(U^*)\cdot
     (\id\o\Delta)(F^{\alpha})\cr}$$
which, if compared to (4.6b), implies that $U$ satisfies the
cocycle condition (4.3).

Vice versa, if $\alpha\in\Aut\A$ is such that $\exists
U\in(\rho^{\alpha}|\rho)$ satisfying the cocycle condition then
define the extension $\hat\alpha$ as follows:
$$\hat\alpha(AF_{\xi}):=\alpha(A)\cdot (\id\o\xi)(UF)\equiv
\alpha(A)F^{\alpha}_{\xi}$$
where we have introduced $F^{\alpha}=UF$. Now one can verify
easily that this is indeed a $^*$-automorphism. Also, $\hat\alpha$
commutes with $\gamma$, since
$$\eqalign{
(\hat\alpha\circ\gamma_X\ \o\ \id)(F)\ &=\ F^{\alpha}(\onne\o
X)\cr
(\gamma_X\circ\hat\alpha\ \o\ \id)(F)\ &=\ UF(\onne\o X)\cr
}$$

The possible ambiguity of the translation $\hat\alpha$ lies in the
ambiguity of choosing a $U\in\Iso(\rho^{\alpha}|\rho)$. If $U'$ is
an other isomorphism then $U^*U'$ is a selfintertwiner of $\rho$
and satisfies the cocycle condition, too. Therefore $U^*U'=\onne\o
g$ with $g\in\hbox{Center}\,\G$ grouplike: $\Delta(g)=g\o g$.
\hfill{\it Q.e.d.}
\smallskip
It follows from this Theorem that the field algebras $\F_{\reven}$
coincide for all $i\in\Z$. Similarly the $\F_{\rodd}$'s coincide
for all $i$. These two field algebras will be denoted respectively
as $\Feven$ and $\Fodd$. They provide the simplest examples of
complete covariant irreducible field algebra extensions of $\A$.
They demonstrate also that such extensions are not unique, since
by Theorem 4.1 the cocycle $R$ of (3.23) ought to be
a coboundary which is not the case for the $R$-matrix of the
double (even for the simplest double $\D(Z(2))=Z(2)\times
Z(2)$).

In order to study the commutation relations of fields at spacelike
separation introduce the following notation. For $\rho=\reven$ let
$F_{2i,2i+1}$ be the master field of $\Fr$. Choose a unitary
cocycle $U\in\Iso(\rho^{\alpha}|\rho)$ and denote it by
$U_{2i-1,2i}$. Let $\hat\alpha$ be the translation automorphism of
$\Feven$ associated to $U$. Then define $F_{2j,2j+1}\in\Feven\o\G$
and $U_{2j-1,2j}\in\A\o\G$ for $j\in\Z$ by the recursions
$$\eqalignno{
  F_{2j,2j+1}\ &=\ U_{2j+1,2j+2}\,F_{2j+2,2j+3}&(4.7)\cr
  U_{2j+1,2j+2}\ &=\ (\alpha\o\idG)(U_{2j-1,2j})&(4.8)\cr}$$
Analogously, for $\rho=\rodd$ one can define the fields
$F_{2j-1,2j}\in\Fodd\o\G$ and the associated charge transporters
$U_{2j,2j+1}\in(\rho_{2j-1,2j}|\rho_{2j+1,2j+2})$.
The following commutation relations can be obtained
$$\eqalignno{
F_{2j,2j+1}^{01}F_{2k,2k+1}^{02}\ &=\ \left\{
   \eqalign{&F_{2k,2k+1}^{02}F_{2j,2j+1}^{01}\cdot(\onne\o
             R^{12})\quad j>k\cr
            &F_{2k,2k+1}^{02}F_{2j,2j+1}^{01}\cdot(\onne\o
             R^{21*})\quad j<k\cr
            }\right.&(4.9a)\cr
F_{2j-1,2j}^{01}F_{2k-1,2k}^{02}\ &=\ \left\{
   \eqalign{&F_{2k-1,2k}^{02}F_{2j-1,2j}^{01}\cdot(\onne\o
             R^{21})\quad j>k\cr
            &F_{2k-1,2k}^{02}F_{2j-1,2j}^{01}\cdot(\onne\o
             R^{12*})\quad j<k\cr
            }\right.&(4.9b)\cr
}$$
Notice that these commutation relations are independent of the
choice of the translation $\hat\alpha$. If (4.9) hold true for
the choice $U_{i,i+1}=(\onne\o s^{-1})T_iT_{i+1}$, $i\in\Z$,
(cf. (3.25)) then they hold true for $U$ replaced by $(\onne\o
g)U$ for any central grouplike unitary $g\in\G$.
\medskip
{\sl 4.4. The "enveloping algebra" of field algebras}
\smallskip
If we want to have operators $Q(X),\ X\in\G$ implementing the
action $\gamma$ of the double on a field algebra $\Fr$ we are lead
to define the crossed product $\B_{\rho}=\Fr\cros\G$ generated by
$F\in\Fr$ and new elements $Q(X)$, $X\in\G$ satisfying
$$\eqalignno{
Q(X)Q(Y)&=Q(XY)&(4.10a)\cr
Q(X^*)&=Q(X)^*&(4.10b)\cr
Q(X)A&=AQ(X)&(4.10c)\cr
Q(X)F_{\xi}&=F_{X_{(1)}\to\xi}Q(X_{(2)})&(4.10d)\cr
}$$
for $X\in\G$, $A\in\A$, and $\xi\in\hat\G$.
Hence $\B_{\rho}$ is the linear span of elements of the
form $AF_{\xi}Q(X)$. In terms of the master field $F$ the
implementation relation (4.10d) takes the form
$$(Q(X)\o\one)F=F\cdot(Q\o\idG)\circ\Delta^{op}(X)\,.\eqno(4.11)$$
$\B_{\rho}$ can also be viewed as the crossed product
$A\cros(\hat\G\cros\G)$ with the simple algebra $\hat\G\cros\G$.
It turns out that $\B_{\rho}$ is actually independent of the
comodule action $\rho$. More precisely,
anticipating the result of section 5, that the special
amplimorphism $\rho_{i,i+1}\colon\A\to\A\o\G$, where $\G$ is the
double $\D(H)$, is universal in the whole category $\Amp\A$, we
can say that $\B_{\rho}$ contains all field algebra extensions of
$\A$ and is independent of $\rho$.
Therefore this
algebra will be called the enveloping algebra of field algebras
and be denoted by $\B$.

In order to prove independence of $\B$ on $\rho$ choose an
arbitrary comodule action $\rho'$ which is equivalent to $\rho$ as
an amplimorphism. Then they are cocycle equivalent as comodule
algebra actions by Proposition 3.6 and we may choose a cocycle
equivalence $(U,u)$
from $(\rho,\Delta)$ to $(\rho',\Delta')$. Let $F\in\F_{\rho}\o\G
\subset\B\o\G$ be the master field of $\rho$ and construct the
unitary $F':=UF(Q\o\id)(u^{*op})$. A straightforward
but lengthy calculation shows that $F'$ satisfies the defining
relations of the master field of $\rho'$:
$$\eqalign{
F'^{01}F'^{02}&=(\id\o\Delta')(F')\cr
F'(A\o\one)&=\rho'(A)F'\cr
F'^*&=(\id\o S')(F')\cr
}$$
with the primed structure maps on the RHS refering to the
twisted double
$(\G,\Delta'=\Ad_u\circ\Delta,\varepsilon,S'=\Ad_v\circ S)$,
where $v=u_1S(u_2)$.

Now we can formulate our main result on the classification of field
algebras.
\smallskip
{\bf Theorem 4.4. } {\sl Equivalence classes of complete
irreducible field algebra
extensions of the observable algebra $\A$ of the Hopf spin model
are in one-to-one correspondence with cohomology classes of
unitary cocycles (3.17). All such field algebra extensions are,
up to equivalence,
crossed products with respect to some coaction of $\G$ on $\A$.
If $\F$ is one complete irreducuble field
algebra then the crossed product $\B=\F\cros\G$ is independent of
the choice of $\F$ and contains all complete irreducible field
algebras as unital $^*$-subalgebras. The translation automorphism
$\alpha$ extends to $\B$ in such a way that its restriction to any
one of the field algebras $\F$ in $\B$ is a translation
$\hat\alpha$ in the sense of Theorem 4.3. Therefore all complete irreducible
field algebras are translation covariant. }

The proof of the theorem will use the following two lemmas.

{\bf Lemma 4.5. } {\sl For the special comodule algebra action
$(\rho,\Delta)=$\break$=(\reven,\Del)$ there exists an
embedding $\Lambda\colon\G\to\A$ of the
double as a $C^*$-algebra into the observable algebra such that
$$\rho\circ\Lambda=(\Lambda\o\id)\circ\Delta\ .\eqno(4.12)$$
}

{\it Proof}: Define $\Lambda(\D(a)):=A_{2i}(a)$ for $a\in H$ and
$\Lambda(\D(\varphi)):=
A_{2i-1}(\varphi_{(2)})$ $A_{2i+1}(\varphi_{(1)})$
for $\varphi\in\hat H$
and verify by straightforward calculation that $\Lambda(\D(a))$
and $\Lambda(\D(\varphi))$ satisfy the defining relations (B.1)
of the double and also the relations
$\reven\circ\Lambda(X)=(\Lambda\o\id)\circ\Del(X)$ for the
generators $X=\D(a)$ and $=\D(\varphi)$.
\hfill{\it Q.e.d.}
\smallskip
{\bf Lemma 4.6. } {\sl Let $(\rho,\Delta)$ be a comodule algebra
action
and $u\in\G\o\G$ be a unitary $\Delta$-cocycle, i.e. satisfies
(3.17). Then there exists a unitary $U\in\A\o\G$ such that the
pair $(U,u)$ is a cocycle equivalence, i.e. (3.16c) holds.}

{\it Proof}: We need to consider only the case
$(\rho,\Delta)=(\reven,\Del)$. For general $(\rho,\Delta)$
the statement follows from the fact that cocycle equivalences can
be composed.

Let $\Lambda$ be the embedding of $\G$ associated to
$(\rho,\Delta)$ in the sense of Lemma 4.5. Then define
$$U\ :=\ (\Lambda\ \o\ \id)(u)\ ,\eqno(4.13)$$
which gives
$$\eqalign{
U\times U&=(U\o\one)\cdot(\rho\o\id)(U)=(\Lambda(u_1)\o u_2\o\one)
           \cdot(\rho\circ\Lambda(u_1)\o u_2)=\cr
         &=(\Lambda\o\id\o\id)\left(\left(u\o\one\right)\cdot
           \left(\Delta\o\id\right)\left(u\right)\right)=\cr
         &=(\Lambda\o\id\o\id)\left(\left(\one\o u\right)\cdot
           \left(\id\o\Delta\right)\left(u\right)\right)=\cr
         &=(\onne\o u)\cdot(\id\o\Delta)(U)\ .\cr
}$$
\hfill{\it Q.e.d.}
\smallskip
{\it Proof of Theorem 4.4.}: The field algebra $\Feven$ associated
to
the special comodule action $\reven$ was shown to be irreducible:
$\A'\cap\Feven=\C\onne$. Anticipating the results of Section 5,
$\Feven$ is also complete, i.e. creates all sectors of $\A$,
because $\reven$ is universal. Hence there exist complete
irreducible field algebras. Let $(\rho,\Delta)$ be a
fixed universal
outer comodule algebra action and $\F$ be the associated complete
irreducible field algebra. Let $\F^{\sharp}$ be any complete irreducible
field algebra. Completeness implies that there exists a unitary
$M\in\F^{\sharp}\o\G$ implementing a universal amplimorphism $\mu$:
$M(A\o\one)=\mu(A)M\,\ A\in\A$. Although $\mu$ is not
necessarily a comodule action there exists an isomorphism of
amplimorphisms $U_0\in\Iso(\rho|\mu)$. Therefore
$F_0:=U_0 M$ is a unitary implementing $\rho$ and satisfying
$$\eqalign{
F_0^{01}F_0^{02}(A\o\one\o\one)&=\rho\times\rho(A)F_0^{01}F_0^{02}\cr
(\id\o\Delta)(F_0)(A\o\one\o\one)&=(\id\o\Delta)\circ\rho(A)
\cdot(\id\o\Delta)(F_0)\cr
}$$
Therefore, by irreducibility of $\F^{\sharp}$,
$$(\id\o\Delta)(F_0^*)\cdot F_0^{01}F_0^{02}\ \in\ \onne\o\G\o\G$$
that is, there exist a unitary $u\in\G\o\G$ such that
$$F_0^{01}F_0^{02}\ =\ (\id\o\Delta)(F_0)\cdot(\onne\o u)
\ .\eqno(4.14)$$
Associativity of $\F^{\sharp}$ implies that this $u$ is a
$\Delta$-cocycle. (Equation (4.14)
expresses the fact that $\F^{\sharp}$ is a projective representation
of the field algebra $\F$ with cocycle $u$.) Now use Lemma 4.6
yielding a cocycle equivalence $(U,u)$ for the cocycle $u$.
With this $U$ we can define $F^{\sharp}:=UF_0$ which turns out to be a
master field for the comodule algebra action
$(\rho^{\sharp}=\Ad_U\circ\rho,\Delta^{\sharp}=\Ad_u\circ\Delta)$.
Having found a master field for $(\rho^{\sharp},\Delta^{\sharp})$
within $\F^{\sharp}\o\G$ means that we have constructed a surjective
$^*$-homomorphism $\delta\colon\F_{\rho^{\sharp}}\to\F^{\sharp}$.
In order to prove that $\delta$ is injective use assumption III that
there exists  a conditional expectation ${\cal E}^{\sharp}\colon
\F^{\sharp}\to\A$ with finite index. For then ${\cal E}:=
{\cal E}^{\sharp}\circ\delta$
is a conditional expectation from the crossed product $\F_{\rho^{\sharp}}$
onto $\A$ and hence by irreducibility of $\A\subset\F_{\rho^{\sharp}}$ it
must coincide with $\gamma_h$ of subsection 4.1. Now the existence of a
quasibasis for ${\cal E}=\gamma_h$ shows that an ideal in $\F_{\rho^{\sharp}}$
such as $\hbox{Ker}\,\delta$ which is annihilated by ${\cal E}$ (${\cal E}
(\hbox{Ker}\,\delta)=\{0\}$) must necessarily be zero:
 $$\hbox{Ker}\,\delta=
\sum_s F_{\eta_s}{\cal E}(F_{\eta_s}^*\hbox{Ker}\,\delta)=\{0\}\,.$$
In this way we have shown that $\F^{\sharp}$ is a crossed product and
have determined a map $\F^{\sharp}\mapsto u$ from the set of complete
irreducible field algebras to the set of $\Delta$-cocycles, for a
fixed $\Delta$. This map is obviously surjective and by Theorem
4.1 equivalence classes of $\F^{\sharp}$-s are mapped to
coboundary equivalence classes of $u$-s.

That $\B$ contains all crossed products $\A\cros\G$ have
already been shown before the formulation of the Theorem.
It remained to prove translation covariance of all complete
irreducible field algebras. Since $\Feven$ is translation
covariant, we have an equivalence
$(U,\one)\in\left(\left(\rho^{\alpha},\Delta\right)|
\left(\rho,\Delta\right)\right)$ where $\rho=\reven$ and
$\Delta=\Del$. Let $(\rho',\Delta')$ be an arbitrary comodule
algebra action. Choose a cocycle equivalence
$(V,u)\in\left(\left(\rho',\Delta'\right)|
\left(\rho,\Delta\right)\right)$.
Hence the composition of cocycle equivalences
$$\eqalign{
((\alpha\o\id)(V),u)\cdot(U,\one)\cdot(V^*,u^*)\ &=:\ (U',\one)\
\in\cr
&\in\left(\left(\rho'^{\alpha},\Delta'\right)|
\left(\rho',\Delta'\right)\right)\cr
}$$
is an equivalence, showing that $(\rho',\Delta')$ is
$\alpha$-covariant. Therefore the corresponding crossed product
$\F_{\rho'}$ is also $\alpha$-covariant.
\hfill{\it Q.e.d.}

\bigskip
{\bf 5. Universality of $\rho$}
\medskip
{\sl 5.1. Compressibility of the net $\A$}
\medskip
Here we derive a property of the net $\{\A(I)\}$
which will enable us to prove that the
amplimorphisms constructed in subsection 3.3 actually exhaust all
possible (localized) amplimorphisms up to equivalence.

Let $J\in\I$ be a non-empty interval of length $|J|=$even.
Define a subset $\I_J$ of the set of intervals $\I$ by
$$\I_J\ :=\ \{I\in\I\ |\ \hbox{either}\ I\subset J'\ \hbox{or}\ I\supset J\}
$$
and a surjective map $\sigma_J\colon\I_J\to\I$ that can be described as
the map arising when we discard $J$ from the
chain and reunite the two remaining parts again:
$$\sigma_J(I)\ :=\ \cases{
I&if $I\subset J'_-$\cr
I-|J|&if $I\subset J'_+$\cr
(I\cap J_-^c)\cup(I\cap J_+^c\ -|J|)&if $I\supset J$\cr
}\eqno(5.1)$$
where $J^c_{\pm}$ denote the two components
of the complement $J^c=\Z\setminus J$. $\sigma_J$ is actually a bijection
since it is the lift to intervals of a bijective map from $\Z\setminus J$
to $\Z$.
\smallskip
{\sl Definition 5.1.}: The relative net over $J$ is defined by
$$\A_J(I)\ =\ \A(\sigma^{-1}_J(I))\ \cap\ \A(J)'\qquad I\in\I$$
providing a net structure for the relative commutant $\A(J)'\cap\A$.
\smallskip
{\bf Theorem 5.2.} ({\it compressibility of the Hopf spin net})
{\sl For any non-empty interval $J$ of even length the relative net
 $\{\A_J(I)\}$ over $J$ is isomorphic to the original. That is there
exists a $^*$-isomorphism $\kappa_J\colon\A\cap\A(J)'\to\A$ such that
$$\kappa_J(\A_J(I))\ =\ \A(I),\qquad I\in\I\,.\eqno(5.2)$$}
{\it Proof}: At first we point out that the algebras $\A_J(I)$ and $\A(I)$
are isomorphic for each $I$. If $\sigma_J^{-1}(I)\subset J'$ then this is
a trivial consequence of translation covariance of the net and the fact that
$|J|=$even. If $\sigma_J^{-1}(I)\supset J$ then the isomorphy follows
from compairing the two inclusions $\A(J)\subset\A(\sigma_J^{-1}(I))$ and
$\C\cdot\onne\subset\A(I)$. By the crossed product structure of the Hopf spin
net and by simplicity of both $\A(J)$ and $\C\cdot\onne$ the two relative
commutants $\A(J)'\cap\A(\sigma_J^{-1}(I))$ and $\C\cdot\onne'\cap\A(I)$
can be represented by path algebras of one and the same inclusion graph,
and therefore are isomorphic.

The proof that a global isomorphism $\kappa_J$ exists will be carried out in
two steps. At first one constructs $\kappa_J$ for $J$ of length $|J|=2$.
This is a technically involved calculation and will be presented afterwards.
Then one shows that the powers of $\kappa_J$ with $|J|=2$ yield net
isomorphisms $\A_J(I)\to\A(I)$ for arbitrary even length intervals $J$.

Now we proceed by assuming that the theorem is proven for $|J|=2$. Construct
the nested sequence of even length intervals $J_1=\{i,i+1\},\dots, J_n
=\{i,\dots,i+2n-1\}$. Then the isomorphism $\kappa_{J_1}$ exists.
Notice that $\sigma_{J_1}(J_{n+1})=J_n$ and that for all $n<m$
$$\sigma_{J_n}^{-1}\ \circ\ \sigma^{-1}_{\sigma_{J_n}(J_m)}\ =
  \ \sigma^{-1}_{J_m}\ .$$
Therefore $\sigma^{-1}_{J_1}\circ\sigma^{-1}_{J_n}=\sigma^{-1}_{J_{n+1}}$.
It follows that
$$\eqalignno{
\A_{J_{n+1}}(I)&=\A\left(\sigma^{-1}_{J_1}\circ\sigma^{-1}_{J_n}(I)\right)
\ \cap\ \A(J_{n+1})'\cr
&\subset
\A\left(\sigma^{-1}_{J_1}\circ\sigma^{-1}_{J_n}(I)\right)
\ \cap\ \A(J_{1})'\ =\ \A_{J_1}\left(\sigma^{-1}_{J_n}(I)\right)&\cr
\kappa_{J_1}\left(\A_{J_{n+1}}(I)\right)\ &\subset\
\A\left(\sigma^{-1}_{J_n}(I)\right)\qquad \forall I\in\I\,.&(5.3)\cr
}$$
On the other hand if $A\in\A_{J_{n+1}}(I)$ and $B\in\A(J_{n+1})$ then
$[A,B]=0$ and especially for $B\in\A(J_{n+1})\cap\A(J_1)'$ in which latter
case $[\kappa_{J_1}(A),\kappa_{J_1}(B)]=0$ follows. Since
$$\kappa_{J_1}\left(\A(J_{n+1})\cap\A(J_1)'\right)=
  \kappa_{J_1}\left(\A\left(\sigma_{J_1}^{-1}(J_n)\right)\cap\A(J_1)'\right)
=\A(J_n)\,,$$
we obtain that
$$\kappa_{J_1}\left(\A_{J_{n+1}}(I)\right)\ \subset\ \A(J_n)'\ .\eqno(5.4)$$
Finally (5.3) and (5.4) together with Definition 5.1 imply that
$$\kappa_{J_1}\left(\A_{J_{n+1}}(I)\right)\ \subset\
\A_{J_n}(I)\qquad I\in\I\ .
\eqno(5.5)$$
Now (5.5) for $n=0,1,\dots,m-1$ implies that the $m$-th power of $\kappa_{J_1}$
yields a $C^*$-inclusion
$$\kappa_{J_1}^m\left(\A_{J_m}(I)\right)\ \subset\ \A(I)\qquad I\in\I$$
between isomorphic $C^*$-algebras and therefore is necessarily an isomorphism.
Hence $\kappa_{J_m}:=\kappa_{J_1}^m$ is a net isomorphism required by the
Theorem.

The case $|J|=2$: Let us assume $J=\{2l,2l+1\}$.
(The case $J=\{2l-1,2l\}$ can be handled analogously.) By additivity of
the net it is enough to define $\kappa_J$ on the 1-point algebras
$\A_J(I)$,$|I|=1$. If $I=\{i\}$
and $i<2l-1$ or $i>2l$ then $\A_J(I)=\A(I)$ or $\A_J(I)=\A(I+2)$, respectively,
and we may define $\kappa_J$
to be the restriction of $\idA$, respectively $\alpha$ onto
$\A_J(I)$. The only non-trivial cases are
$I=\{2l-1\}$ and $I=\{2l\}$. To handle them the
following Lemma will be useful.

{\bf Lemma 5.3. } {\sl Let $\varphi\in\hat H$, $a\in H$, and the
indices $i,k$ run over $1,\dots,N=\hbox{dim}H$. We set
$D^{ik}(\varphi)=\langle b^i,\varphi\beta^k\rangle$ and
$D^{ik}(a)=\langle b^i,a\to \beta^k\rangle$
 for the left regular representation of $\hat H$ and $H$,
respectively, with a fixed pair of dual bases $\{\beta^i\}$ and
$\{b^i\}$. If furthermore $E^{ik}$ denote
$C^*$-matrix units for the simple algebra $\A_{2l,2l+1}$ then
$$\eqalign{
 E^{ik}(\varphi)&:={1\over N}\sum_{j,m}\
E^{ij}A_{2l-1}(\varphi_{(1)})
 D^{jm}(\varphi_{(2)})E^{mk}\cr
 E^{ik}(a)&:={1\over N}\sum_{j,m}\
E^{ij} D^{jm}(a_{(1)})A_{2l+2}(a_{(2)})E^{mk}\cr}$$
provide us with bases of the 3-point algebras
$\A(\{2l-1,2l,2l+1\})$ and $\A(\{2l,2l+1,2l+2\})$, respectively.
They satisfy
$$E^{ik}(\varphi)E^{mn}(\psi)=\delta_{k,m}E^{in}(\varphi\psi)\,,
\quad
E^{ik}(a)E^{mn}(b)=\delta_{k,m}E^{in}(ab)\,,$$
hence providing isomorphisms of the two 3-point algebras with
$\hat H\o \End H$ and $H\o\End H$ respectively.}
\smallskip
\noindent The proof of this Lemma is a rather elementary exercise
with Hopf algebra identities, so  will be omitted.

Continuing with the proof of the Theorem notice that a
consequence of this Lemma is that the partial traces
$E(\varphi)=\sum_k E^{kk}(\varphi)$ and $E(a)=\sum_kE^{kk}(a)$
yield the relative commutants of $\A(J)$ within the two
3-point algebras:
$$\eqalign{
\{E(\varphi)\,|\,\varphi\in\hat H\}
&=\A(\{2l,2l+1\})'\cap\A(\{2l-1,2l,2l+1\})\equiv\A_J(\{2l-1\})\cr
\{E(a)\,|\,a\in H\}
&=\A(\{2l,2l+1\})'\cap\A(\{2l,2l+1,2l+2\})\equiv\A_J(\{2l\})\cr}$$
Therefore $\kappa_J$ is defined by setting $\kappa_J(E(\varphi)):=
A_{2l-1}(\varphi)$ and $\kappa_J(E(a)):=A_{2l}(a)$.

In order to show that $\kappa_J$ is a$^*$-homomorphism we need to check
only the commutation relations
between the neighbouring 1-point algebras. There are 3 non-trivial cases:
the relations between $\A_J(\{2l-2\})$ and $\A_J(\{2l-1\})$,
between $\A_J(\{2l-1\})$ and $\A_J(\{2l\})$ and between
$\A_J(\{2l\})$ and $\A_J(\{2l+1\})$. The first and the third of
these can be checked rather easily: For example
$$\eqalign{
E(\varphi)A_{2l-2}(a)&={1\over N}\sum_{ijk}\ E^{ij}
A_{2l-1}(\varphi_{(1)})A_{2l-2}(a) D^{jk}(\varphi_{(2)})E^{ki}=\cr
&={1\over N}\sum_{ijk}\ E^{ij}A_{2l-2}(\varphi_{(1)}\to a)
A_{2l-1}(\varphi_{(2)})D^{jk}(\varphi_{(3)})E^{ki}\cr
&=A_{2l-2}(\varphi_{(1)}\to a)E(\varphi_{(2)})\cr}$$
The commutation relations between $\A_J(\{2l-1\})$ and
$\A_J(\{2l\})$
can be obtained as follows. At first one computes the commutation
relations
$$E^{ik}A_{2l-1}(\varphi)=A_{2l-1}(\varphi_{(1)})E^{i'k'}\cdot
D^{ii'}(\varphi_{(2)})D^{k'k}(S(\varphi_{(3)}))$$
using the explicit representation of the matrix units [N]:
$$E^{ik}=\sum_j\
A_{2l}(b^iS^{-1}(b^j))A_{2l+1}(\beta^j\beta^k)\,.\eqno(5.6)$$
Then one obtains
$$\eqalign{
E(a)E(\varphi)&={1\over N^2}\sum_{ijkmn}\
E^{ij} D^{jk}(a_{(1)})
A(a_{(2)})E^{km}A(\varphi_{(1)})D^{mn}(\varphi_{(2)})E^{ni}\cr
&={1\over N^2}\sum_{ijkmn}\sum_{k'm'}\ E^{ij}
D^{jk}(a_{(1)})
A(a_{(2)})A(\varphi_{(1)})D^{kk'}(\varphi_{(2)})E^{k'm'}\cr
&\qquad
D^{m'm}(S(\varphi_{(3)}))D^{mn}(\varphi_{(4)})E^{ni}\cr
&={1\over N}\sum_{ijm}\ E^{ij}A(\varphi_{(1)})\cdot\sum_k
D^{jk}(a_{(1)})D^{km}(\varphi_{(2)})\cdot
A(a_{(2)})E^{mi}\cr
&={1\over N}\sum_{ijm}\
E^{ij}A(\varphi_{(1)})\cdot\sum_k
D^{jk}(\varphi_{(2)})\langle
\varphi_{(3)},a_{(1)}\rangle D^{km}(a_{(2)})\cdot
A(a_{(3)})E^{mi}\cr
}\eqno(5.7)$$
where in the last line we used the identity
$$\sum_k\ D^{jk}(a)D^{km}(\varphi)=\sum_k\ D^{jk}(\varphi_{(1)})
\langle\varphi_{(2)},a_{(1)}\rangle D^{km}(a_{(2)})$$
expressing the fact that $D$ is a representation of the whole Weil
algebra $\W(H)=\hat H\cros H$. Now using the
commutation relations
$$E^{ik}A_{2l+2}(a)=A_{2l+2}(a_{(3)})E^{i'k'}\cdot
D^{ii'}(a_{(2)})D^{k'k}(S(a_{(1)}))$$
we can compute the expression
 $$\eqalign{
&E(\varphi_{(1)})\langle\varphi_{(2)},a_{(1)}\rangle
E(a_{(2)})=\cr
&={1\over N^2}\sum_{ijkmn}\
E^{ij}A(\varphi_{(1)})D^{jk}(\varphi_{(2)})E^{km}
\langle\varphi_{(3)},a_{(1)}\rangle D^{mn}(a_{(2)})
A(a_{(3)})E^{ni}=\cr
&={1\over N^2}\sum_{ijkmn}\sum_{k'm'}\ E^{ij}A(\varphi_{(1)})
D^{jk}(\varphi_{(2)})\langle\varphi_{(3)},a_{(1)}\rangle
A(a_{(5)})E^{k'm'}D^{kk'}(a_{(4)})\cr
&\qquad D^{m'm}(S(a_{(3)}))D^{mn}(a_{(2)})
E^{ni}=\cr
&={1\over N}\sum_{ijkm}\
E^{ij}A(\varphi_{(1)})D^{jk}(\varphi_{(2)})
\langle\varphi_{(3)},a_{(1)}\rangle D^{km}(a_{(2)})A(a_{(3)})
E^{mi}\cr}$$
which, when compared to (5.7), yields finally
$$E(a)E(\varphi)=E(\varphi_{(1)})\langle\varphi_{(2)},a_{(1)}
\rangle E(a_{(2)})\,.\eqno(5.8)$$
In this way we have constructed a $^*$-homomorphism $\kappa_J$ mapping
$\A_J(I)$ into $\A(I)$ for all $I$. To see that it is actually an isomorphism
one simply constructs its inverse by defining it on the 1-point algebras
in the obvious way.\hfill{\it Q.e.d.}
\medskip

{\sl 5.2. Compressibility of the amplimorphisms}
\smallskip
The following theorem will show that compressibility of the chain
$\A$ has very
strong consequences on the structure of amplimorphisms of $\A$.
\smallskip
{\bf Theorem 5.4. } ({\it Compressibility of the amplimorphisms})
{\sl Let $\mu$ be a localized amplimorphism of
the Hopf spin chain. Then $\mu$ is equivalent to an amplimorphism
$\mu_0$ that is localized within an interval of length 2.}

{\it Proof}: Let $\mu\colon\A\to\A\o\End V$ be a localized
amplimorphism. Choose an interval
$I$ of length $|I|=$even and $|I|\geq 4$ such that $\mu$ is
localized within $I$. Define the interior of $I$ by $\Int I:=
(I^c)'$.
Then by Haag duality $\mu(\A(\Int I))\subset
\A(\Int I)\o\End V$ and by the split property $\A(\Int I)$ is
simple. Since any amplimorphism of a simple (finite dimensional)
algebra is inner, there exists a unitary $U\in\A(\Int I)\o\End V$
such that
$$\mu(A)\ =\ U(A\o\one_V)U^*\,,\qquad A\in\A(\Int
I)\,.\eqno(5.9)$$
Let $\underline{\mu}$ denote the amplimorphism
$\Ad_{U^*}\circ\mu$, then $\underline{\mu}$ acts as the amplified
identity on $\A(\Int I)$. It follows that for all interval
$J\supset \Int I$ we have
$$\underline{\mu}(\A(J)\cap\A(\Int I)')\ \subset\
  (\A(J)\cap\A(\Int I)')\ \o\ \End V$$
Therefore $\underline{\mu}$ can be restricted to the relative net
$\A_{\Int I}=\A\cap\A(\Int I)'$ to yield an amplimorphism $\mu_0$
localized on the interval $I_0=\sigma_{\Int I}(I)$ (cf. Def.5.1) of length
2 and satisfying
$$\mu_0\circ\kappa^{-1}\ =\
(\kappa^{-1}\o\id)\circ\underline{\mu}$$
where $\kappa$ denotes the compressing isomorphism
$\kappa\colon\A_{\Int I}\to\A$ constructed in Theorem 5.2.

The map $\underline{\mu}\mapsto\mu_0$, from the set of "smeared"
amplimorphisms $\underline{\mu}$ to the 2-point amplimorphisms
$\mu_0$, can be inverted. As a matter of fact, using additivity of
the net, we can define $\underline{\mu}$ on the subalgebra
$\A(\Int I)'$ as $(\kappa\o\id)\circ\mu_0\circ\kappa^{-1}$ and
extend it to $\A$ by letting it to act on $\A(\Int I)$ as the
trivial amplification.

Therefore there is a one-to-one correspondence between
amplimorphisms localized within $I$ and smeared over $\Int I$
and the 2-point amplimorphisms localized within $I_0$.
Since the intertwiners $T\in(\underline{\mu}|\underline{\nu})$
between two smeared amplimorphisms belong to the commutant of both
$\A(I')$ and $\A(\Int I)$, they are scalars. Therefore they are
mapped bijectively onto the intertwiner space $(\mu_0|\nu_0)$.
This proves that the category of amplimorphisms localized within
$I$ is equivalent to the category of amplimorphisms localized
within $I_0$, the latter one being a subcategory of the former.
In other words all the charges that can be created on a finite
interval can also be created on an interval of length 2.
\hfill{\it Q.e.d.}
\smallskip
An other striking consequence of compressibility is that
all amplimorphisms (i.e. localized $C^*$-maps from $\A$ to
$\A\o\End V$) are transportable in the following very strong
sense: If $\mu$ is localized within some $I\in\I$ and $J\in\I$ is
an arbitrary interval of length at least 2 then there exists a
$\nu$ localized within $J$ that is equivalent to $\mu$.
The proof goes as follows. At first we remark that there is a left
analogue of the shrinking map $\sigma_J$ of (5.1) that uses translation
on the left hand side of $J$. Combining the left and right $\sigma_J$-s
we have a slightly different relative net but can construct a compression
isomorphism $\kappa_J$ as in Theorem 5.2. Consequently Theorem 5.4 will
imply that an amplimorphism localized within $\{i,\dots,i+2n+1\}$, $i\in\Z,
n\in\N$, is equivalent to an amplimorphism localized within any 2-point
subintervals of the form $\{i+2m,i+2m+1\}$.
Now the general transportability property easily follows.
Choose an interval $K$ of length even which
contains both $I$ and $J$. Let $\underline{\mu}$ be the smearing of $\mu$
over $K$ and let $\mu_0$ be the 2-point amplimorphism localized
within $K_0$ arising as the compression of $\underline{\mu}$.
Obviously $\mu\sim\underline{\mu}\sim\mu_0$. By the above remark $K$ and
$K_0$ can be chosen in such a way that $K_0\subset J$. Hence $\mu$ is
equivalent to an amplimorphism localized within $J$.

In order to be able to conclude that all superselection sectors of
the Hopf spin model arise from the application of the special
comodule action $\rho_{i,i+1}$ of subsection 3.3 (i.e. that the
special comodule action is universal in the category $\Amp\A$) we
only have
to find the general form of an amplimorphism localized within a
2-point interval $\{i,i+1\}$. This will be done in the next
subsection.

 \medskip
{\sl 5.3. The 2-point amplimorphisms}
\smallskip
{\bf Proposition 5.4. } {\sl Let $\mu\colon\A\to\A\o M_n$ be an
amplimorphism localized within an interval $I$ of length 2 and let
$\rho$ denote
the comodule action $\rho_I$ defined in (3.18) which is localized
within the same interval.
Then there exists a non-zero intertwiner $T\in(\rho|\mu)$.}

{\it Proof}: We restrict ourselves to the case $I=\{2i,2i+1\}$.
{}From Haag duality of the net it follows that $\mu(\A_{2i})\subset
\A_{2i}\o M_n$ and $\mu(\A_{2i+1})\subset\A_{2i+1}\o M_n$. Hence
we may define the $^*$-algebra maps
$\mur\colon\hat H\to M_n\o\hat H$ and $\mul\colon H\to H\o M_n$
by
$$\mur(\varphi)=\tau_{01}\circ\mu(A_{2i+1}(\varphi))\qquad
  \mul(a)=\mu(A_{2i}(a))\eqno(5.10)$$
Using commutation relations with $\A_{2i+2}$ and $\A_{2i-1}$
repectively, we can write
\smallskip
{\settabs 2\columns
\+$(I_n\o a)\mur(\varphi)=\mur(a_{(1)}\to\varphi)(I_n\o a_{(2)})$,
 &$\mul(a)(\varphi\o I_n)=(\varphi_{(1)}\o I_n)\mul(a\leftarrow
      \varphi_{(2)})$,\cr
\+$(\id_{M_n}\o a\to)(\mur(\varphi))=\mur(a\to\varphi)$,
 &$(\leftarrow \varphi\o\id_{M_n})(\mul(a))=\mul(a\leftarrow
      \varphi)$,\cr
\+$(\id_{M_n}\o\Delta_{\hat H})\circ\mur=(\mur\o\id_{\hat H})
     \circ\Delta_{\hat H}$,
 &$(\Delta_H\o\id_{M_n})\circ\mul=(\id_H\o\mul)\circ\Delta_H$,\cr
}
\smallskip
\noindent
Applying $\id_{M_n}\o\varepsilon_{\hat H}\o\id_{\hat H}$ and
$\id_H\o\varepsilon_H\o\id_{M_n}$ respectively
$$\mur(\varphi)=\mure(\varphi_{(1)})\o\varphi_{(2)}\,,
\qquad\mul(a)=a_{(1)}\o\mule(a_{(2)})$$
with $\mure\colon\hat H\to M_n$, $\mule\colon H\to M_n$ being
unital $C^*$-maps. Therefore
$$\eqalign{
\mu(A_{2i+1}(\varphi))&=A_{2i+1}(\varphi_{(2)})\ \o\
\mure(\varphi_{(1)})\quad\varphi\in\hat H\cr
\mu(A_{2i}(a))&=A_{2i}(a_{(1)})\ \o\
\mule(a_{(2)})\quad a\in H\cr}
\eqno(5.11)$$
It remained to investigate the mutual commutations of
$\mu(\A_{2i})$ and $\mu(\A_{2i+1})$.
$$\eqalign{
&\mu(A_{2i+1}(\varphi))\mu(A_{2i}(a))=\cr
&\qquad A_{2i}(a_{(1)})\,\langle a_{(2)},\varphi_{(2)}\rangle\,
        A_{2i+1}(\varphi_{(3)})\ \o\ \mure(\varphi_{(1)})
        \mule(a_{(3)})\cr
&\mu(A_{2i+1}(\varphi)A_{2i}(a))=\cr
&\qquad A_{2i}(a_{(1)})\,\langle a_{(3)},\varphi_{(1)}\rangle\,
        A_{2i+1}(\varphi_{(3)})\ \o\ \mule(a_{(2)})
        \mure(\varphi_{(2)})\cr
        }$$
Multiplying both of these equations by $A_{2i}(S(a_{(0)}))$ from
the left and by\break
$A_{2i+1}(S(\varphi_{(4)}))$ from the right
we obtain
$$\langle a_{(1)},\varphi_{(2)}\rangle \,\mure(\varphi_{(1)})
  \mule(a_{(2)})\ =\
  \langle a_{(2)},\varphi_{(1)}\rangle \,\mule(a_{(1)})
  \mure(\varphi_{(2)})$$
which is but the defining relation (B.1c) of the double $\G$.
Hence
$$\D(a)\D(\varphi)\in\G\quad {\buildrel\theta\over\mapsto}\quad
  \mule(a)\mure(\varphi)\in M_n$$
defines a non-zero $^*$-algebra homomorphism by means of which we
can express $\mu$ as
$$\mu\ =\ (\idA\o \theta)\circ\rho\eqno(5.12)$$
which obviously implies the existence of a non-zero
$T\in(\rho|\mu)$. As a matter of fact, since $\theta\neq 0$, there
exists a non-zero $t\in\Hom(\C^n,V)$ such that $xt=t\theta(x),\
x\in\G$. Then $T=\one\o t$ is the required intertwiner.
\hfill {\it Q.e.d.}

This Proposition together with Theorem 5.4 imply that
amplimorphisms $\reven$ of (3.18) are universal in the whole
category $\Amp\A$. Therefore Theorems 3.1 and 3.10 yield finally
the equivalence of $\Rep\D(H)$  with the category $\Rep\A$ of
all DHR-representations of $\A$. As a byproduct we obtain that
all amplimorphisms are $\alpha$-covariant since $\reven$ was shown
to be $\alpha$-covariant by (3.25--26) which immediately gives
$\Amp^{\alpha}\A=\Amp\A$.

\bigskip
{\bf Appendix A:}\ {\sl Finite dimensional C$^*$-Hopf algebras}
\medskip
There is an extended literature on Hopf algebra theory the
nomenclature of which, however, is by far not unanimous [Sw,
ES, Dr1-2, BaSk]. Therefore we
summarize in this appendix some standard notions in order to fix
our conventions and notations.

A linear space $B$ over $\C$ together with linear maps
$$\eqalign{
\eqalign{m\colon B\o B&\to B\quad \hbox{(multiplication)},\cr
\iota\colon\C&\to B\quad\hbox{(unit)},\cr}
&\eqalign{\Delta\colon B&\to B\o
B\quad\hbox{(comultiplication)},\cr
\varepsilon\colon B&\to\C\quad\hbox{(counit)}\cr}\cr
}$$
is called a {\it bialgebra } and denoted by $B(m,\iota,\Delta,\varepsilon)$
if the following axioms hold:
$$\eqalign{
\eqalign{m\circ(m\o\id)&=m\circ(\id\o m)\,,\ \cr
         m\circ(\iota\o\id)&=m\circ(\id\o \iota)=\id\,,\ \cr
         \varepsilon\circ m&=\varepsilon\o\varepsilon\,,\ \cr}
&\eqalign{(\Delta\o\id)\circ\Delta&=(\id\o\Delta)\circ\Delta\cr
   (\varepsilon\o\id)\circ\Delta&=(\id\o\varepsilon)\circ\Delta
   =\id\cr
         \Delta\circ\iota&=\iota\o\iota\cr}
\cr
\Delta\circ m=(m\o m)\circ\tau_{23}&\circ(\Delta\o\Delta)\cr
}$$
where $\tau_{23}$ denoes the permutation of the tensor factors
2 and 3.
We use Sweedler's notation $\Delta(x)=x_{(1)}\o x_{(2)}$, where
the right hand side is understood as a sum $\sum_i
x_{(1)}^i\o x_{(2)}^i\in B\o B$. For iterated coproducts we
write $x_{(1)}\o x_{(2)}\o x_{(3)}:=\Delta(x_{(1)})\o
x_{(2)}\equiv x_{(1)}\o\Delta(x_{(2)})$, etc.
The image under $\iota$ of the number $1\in\C$ is the unit element of
$B$ and denoted by $\one$. The linear dual $\hat B$ becomes also a
bialgebra by transposing the structural maps
$m,\iota,\Delta,\varepsilon$ by means of the canonical pairing
$\langle\ \ ,\ \ \rangle\colon \hat B\times B\to\C$.

A bialgebra $H(m,\iota,\Delta,\varepsilon)$ is called a {\it Hopf
algebra } $H(m,\iota,S,\Delta,\varepsilon)$ if there exists an
antipode $S\colon H\to H$, i.e. a linear map satisfying
$$
m\circ(S\o\id)\circ\Delta=m\circ(\id\o S)\circ\Delta=
\iota\circ\varepsilon
\eqno(A.1)$$
Using the above notation equ. (A1) takes the form
$S(x_{(1)})x_{(2)}=x_{(1)}S(x_{(2)})=\varepsilon(x)\one$,
which in connection with the coassociativity of $\Delta$ is
often applied in formulas involving iterated coproducts like,
e.g., $x_{(1)}\o x_{(4)}S(x_{(2)})x_{(3)}=x_{(1)}\o x_{(2)}$.
All other properties of the antipode, i.e. $S(xy)=S(y)S(x),\
\Delta\circ S=(S\o S)\circ\Delta_{op}$ and
$\varepsilon\circ S=\varepsilon$, as well as the uniqueness of
$S$ are all consequences of the
axiom (A.1) [Sw].
The dual bialgebra $\hat H$ of $H$ is also a Hopf algebra with the
antipode defined by
$$
\langle S(\varphi),x\rangle:=
\langle\varphi,S(x)\rangle\quad\varphi\in\hat H,\ x\in H\,.
\eqno(A.2)
$$
A $*$-Hopf algebra $H(m,\iota,S,\Delta,\varepsilon,*)$ is a Hopf
algebra $H(m,\iota,S,\Delta,\varepsilon)$ together with an antilinear
involution $^*\colon H\to H$ such that $H(m,\iota,*)$ is a
$*$-algebra and $\Delta$ and $\varepsilon$ are $^*$-algebra
maps.
It follows that $\overline S:=*\circ S\circ *$ is the antipode
in the Hopf algebra
$H_{op}$ (i.e. with opposite muliplication) and therefore
$\overline S=S^{-1}$ [Sw].

The dual of a $*$-Hopf algebra is also a $*$-Hopf algebra with
$^*$-operation defined by
$$
\langle
\varphi^*,x\rangle:=\overline{\langle\varphi,S(x)^*\rangle}\,.
\eqno(A.3)
$$

Let $\A$ be a $*$-algebra and $H$ be a $*$-Hopf algebra. A (Hopf
module) {\it left action}
of $H$ on $\A$ is a linear map $\gamma\colon H\o \A\to \A$
satisfying the following axioms: For $A,B\in \A$, $x,y\in H$
$$
\eqalign{
\gamma_x\circ\gamma_y(A)&=\gamma_{xy}(A)\cr
\gamma_x(AB)&=\gamma_{x_{(1)}}(A)\gamma_{x_{(2)}}(B)\cr
\gamma_x(A^*)&=\gamma_{S(x)^*}(A)^*\cr
}\eqno(A.4)
$$
A {\it right action} of $H$ is a left action of $H_{op}$.
Important examples are the action of $H$ on $\hat H$ and that of
$\hat H$ on $H$ given by the Sweedler's arrows:
$$\eqalignno{
\gamma_x(\varphi)=x\to\varphi&:=\varphi_{(1)}\langle
x,\varphi_{(2)}\rangle&(A.5a)\cr
\gamma_{\varphi}(x)=\varphi\to x&:=
x_{(1)}\langle\varphi,x_{(2)}\rangle&(A.5b)\cr
}$$
A left action is called inner if there exists a *-algebra map
$i:H\to\A$ such that
$\gamma_x(A)=i(x_{(1)})\,A\,i(S(x_{(2)}))$. Left $H$-actions
$\gamma$ are in one-to-one corespondence with right {\it $\hat
H$-coactions} (often denoted by the same symbol)
$\gamma:\A\to\A\o \hat H$ defined by
$$
\gamma(A):=\gamma_{b_i}(A)\o\xi^i,\quad A\in\A
$$
where $\{b_i\} $ is a basis in $H$ and $\{\xi^i\}$ is the dual
basis in $\hat H$ and where for simplicity we assume from now
on $H$ to be finite dimensional. Conversely, we have
$\gamma_x=(\idA\o x)\circ\gamma$. The defining properties of a
coaction are given in equs. (3.11a-e).

Given a left $H$-action (right $\hat H$-coaction) $\gamma$ one
defines the {\it crossed product}
$\A\cros_\gamma H$ as the $\C$-vector space $\A\o H$ with
$*$-algebra structure
$$\eqalignno{
(A\o x)(B\o y) &:= A\gamma_{x_{(1)}}(B)\o x_{(2)}y &(A.6a)\cr
(A\o x )^* &:= (\one_\A\o x^*)(A^*\o\one_H) &(A.6b)\cr
}$$
An important example is the "Weyl algebra" $\W(\hat H):=\hat
H\cros H$, where the crossed product is taken with
respect to the natural left action (A.5a). We have $\W(\hat
H)\cong\End\hat H$ where the isomorphism is given by (see
[N] for a review)
$$
w:\ \psi\o x\mapsto Q^+(\psi)P^+(x)\ .\eqno(A.7)$$
Here we have introduced $Q^+(\psi),\ \psi\in\hat H$ and
$P^+(x),\ x\in H$ as operators in $\End\hat H$ defined on
$\xi\in\hat H$ by
$$\eqalign{
Q^+(\psi)\xi &:= \psi\xi\cr
P^+(x)\xi &:= x\to\xi\cr}$$

Any right $H$-coaction $\beta\,:\A\to\A\o H$ gives rise to a natural
left $H$-action $\gamma$ on $\A\cros_\beta\hat H$
$$
\gamma_x(A\o\psi):=A\o(x\to\psi)\eqno(A.8)$$
The resulting double crossed product $(\A\cros_\beta\hat
H)\cros_\gamma H$ contains $\W(\hat H)\cong\End\hat H$ as the
subalgebra given by $\one_\A\o\psi\o x\cong Q^+(\psi)P^+(x),\
\psi\in\hat H,\ x\in H$. Moreover, by the Takesaki duality
theorem [NaTa] the double crossed product $(\A\cros_\beta\hat H)
\cros_\gamma H$ is canonically isomorphic to $\A\o\End \hat H$. In fact,
defining the representation $L:H\to\End\hat H$ by
$$
L(x)\xi:=\xi\leftarrow
S^{-1}(x)\equiv\langle\xi_{(1)}\,,\,S^{-1}(x)\rangle\xi_{(2)}\eqno(A.9)$$
one easily verifies that $\T:(\A\cros_\beta\hat H)\cros_\gamma H
\to\A\o\End\hat H$
$$\eqalignno{
\T(A\o\one_{\hat H}\o\one_H)&:=(\idA\o L)(\beta(A))&(A.10a)\cr
\T(\one_\A\o\psi\o x)&:=\one_\A\o Q^+(\psi)P^+(x)&(A.10b)\cr
}$$
defines a $*$-algebra map. $\T$ is surjective since $w$ is
surjective and therefore $\one_\A\o\End\hat H\subset\hbox{Im}\T$ and
$$\eqalign{
A\o\one_{\End\hat H}&\equiv A_{(0)}\o L(A_{(1)}S(A_{(2)}))\cr
&=\T(A_{(0)}\o\one_{\hat H }\o\one_H)(\one_\A\o L(S(A_{(1)})))\cr
&\in\hbox{Im}\T\cr
}$$
for all $A\in\A$.
Here we have used the notation
$A_{(0)}\o A_{(1)}=\beta(A)$,
$$
A_{(0)}\o A_{(1)}\o A_{(2)}=(\beta\o\id_H)(\beta(A))\equiv
(\idA\o\Delta)(\beta(A))
$$
(including a summation convention) and the identity
$(\idA\o\varepsilon)\circ\beta=\idA$,
see equs. (3.11d,e).
The inverse of $\T$ is given by
$$\eqalignno{
\T^{-1}(\one_A\o W)&=\one_\A\o w^{-1}(W)&(A.11a)\cr
\T^{-1}(A\o\one_{\End\hat H}) &= A_{(0)}\o w^{-1}(L(S(A_{(1)})))&(A.11b)\cr
}$$
for $W\in\End\hat H$ and $A\in\A$.

\bigskip
A {\it left(right) integral} in $\hat H$
is an element $\chi^L(\chi^R)\in\hat H$ satisfying
$$
\varphi\chi^L=\chi^L\cdot\varepsilon(\varphi)
\qquad\chi^R\varphi=\varepsilon(\varphi)\cdot\chi^R\eqno(A.12a)
$$
for all $\varphi\in\hat H$ or equivalently
$$
\chi^L\rightarrow x=\langle\chi^L,x\rangle\one\,,\qquad
  x\leftarrow\chi^R=\langle\chi^R,x\rangle\one
\eqno(A.12b)
$$
for all $x\in H$. Similarly one defines left(right) integrals
in $H$.

If $H$ is finite dimensional and semisimple then so is
$\hat H$ [LaRa] and in this case they are both {\it unimodular},
i.e. left and right integrals coincide and are all given as
scalar multiples of a unique one dimensional central projection
$$
e_\varepsilon=e_\varepsilon^*=e_\varepsilon^2=S(e_\varepsilon)
\eqno(A.13)$$
which is then called the {\it Haar integral}.

For $\varphi,\psi\in\hat H$ and $h\equiv e_\varepsilon\in H$
the Haar integral define the hermitian form
$$
\langle\varphi|\psi\rangle:=\langle\varphi^*\psi,h\rangle
\eqno(A.14)$$
Then $\langle\cdot|\cdot\rangle$ is nondegenerate [LaSw] and it is
positve definite --- i.e. the Haar integral
$h$ provides a positive state ({\it the Haar "measure"}) on $\hat H$
--- if and only if $\hat H$ is a {\it $C^*$-Hopf
algebra}.
These are the "finite matrix pseudogroups" of [Wo].
They also satisfy $S^2=\id$ and $\Delta(h)=\Delta_{op}(h)$ [Wo].
If $\hat H$ is a finite dimensional $C^*$-Hopf algebra then so is
$H$, since $H\ni x\to P^+(x)\in\End\hat H$ defines a faithful
$*$-representation on the Hilbert space $\Hil\equiv L^2(\hat H,h)$.
Hence finite dimensional $C^*$-Hopf algebras always come in
dual pairs. Any such pair serves as a building block for our Hopf spin
model.

\bigskip

{\bf Appendix B:} \ {\sl The Drinfeld Double}

Here we list the basic properties
of the Drinfeld double $\D(H)$ (also called quantum double)
of a finite dimensional $*$-Hopf
algebra $H$ [Dr1-2, Maj2]. Although most of them are well known in the
literature, the presentation (B.1) by generators and relations
given below seems to be new.

As a $*$-algebra $\D(H)$ is generated by elements $\D(a),\ a\in
H$ and $\D(\varphi),\ \varphi\in\hat H$ subjected to the following
relations:
$$\eqalignno{
\D(a)\D(b)&=\D(ab)
&(B.1a)\cr
\D(\varphi)\D(\psi)&=\D(\varphi \psi)
&(B.1b)\cr
\D(a_{(1)})\,\langle
a_{(2)},\varphi_{(1)}\rangle\,\D(\varphi_{(2)})&=
\D(\varphi_{(1)})\,\langle
\varphi_{(2)},a_{(1)}\rangle\,\D(a_{(2)})
&(B.1c)\cr
\D(a)^*=\D(a^*) &, \D(\varphi)^*=\D(\varphi^*)&(B.1d)\cr
}$$
The relation (B.1c) is equivalent to any one of the following
two relations
$$\eqalignno{
\D(a)\D(\varphi)&=\D(\varphi_{(2)})\D(a_{(2)})\,
  \langle a_{(1)},\varphi_{(3)}\rangle
  \langle S^{-1}(a_{(3)}),\varphi_{(1)}\rangle
&(B.2a)\cr
\D(\varphi)\D(a)&=\D(a_{(2)})\D(\varphi_{(2)})\,
  \langle \varphi_{(1)},a_{(3)}\rangle
  \langle S^{-1}(\varphi_{(3)}),a_{(1)}\rangle
&(B.2b)\cr}$$
These imply that as a linear space $\D(H)\cong H\o\hat H$
and also that as a $*$-algebra $\D(H)$ and $\D(\hat H)$ are
isomorphic. This $*$-algebra will be denoted by $\G$.

The Hopf algebraic structure of $\D(H)$ is given by the following
coproduct, counit, and antipode:
$$\eqalignno{
\Del(\D(a))=\D(a_{(1)})\o \D(a_{(2)}) &\quad
\Del(\D(\varphi))=\D(\varphi_{(2)})\o \D(\varphi_{(1)})
&(B.3a)\cr
\varepsilon_{\D}(\D(a))=\varepsilon(a) &\quad
\varepsilon_{\D}(\D(\varphi))=\varepsilon(\varphi)
&(B.3b)\cr
S_{\D}(\D(a))=\D(S(a)) &\quad
S_{\D}(\D(\varphi))=\D(S^{-1}(\varphi))
&(B.3c)\cr}$$
It is straightforward to check that equs. (B.3) provide a $*$-Hopf algebra
structure on  $\D(H)$. Moreover,  $\D(\hat H)= (\D(H))_{cop}$
(i.e. with opposite coproduct) by (B.3a).

If $H$ and $\hat H$ are $C^*$-Hopf algebras then so is $\D(H)$.
To see this one checks that
$$
\D(h)\D(\chi)=\D(\chi)\D(h)=:h_\D
\eqno(B.4)$$
provides the Haar integral in  $\D(H)$ and that the positivity
of the Haar states $h\in H$ and $\chi\in\hat H$ implies the positvity of the
state $h_\D$ on $\widehat{\D(H)}$ .

The dual $\widehat{\D(H)}$ of $\D(H)$ has been studied by [PoWo].
As a coalgebra it is $\hat \G$ and coincides
with the coalgebra $\widehat{\D(\hat H)}$. The latter one,
however, as an algebra differs
from $\widehat{\D(H)}$ in that the multiplication is replaced
by the opposite multiplication.

\medskip
The remarkable property of the double construction is that it
always yields a {\it quasitriangular} Hopf algebra [Dr1-2].
By definition this means that there exists a unitary $R\in \D(H)\o
\D(H)$ satisfying the hexagonal identities
$R^{13}R^{12}=(\id\o\Delta)(R)$, $R^{13}R^{23}=(\Delta\o\id)(R)$,
and the intertwining property $R\Delta(x)=\Delta^{op}(x)R,\
x\in \D(H)$, where $\Delta^{op}\colon x\mapsto x_{(2)}\o
x_{(1)}$.

If $\{b_A\}$ and
$\{\beta^A\}$ denote bases of $H$ and $\hat H$, respectively,
that are dual to each other, $\langle
\beta^A,b_B\rangle=\delta^A_B$, then
$$
R\equiv R_1\o R_2:=\sum_A\ \D(b_A)\o\D(\beta^A)
\eqno(B.5)$$
is independent of the choice of the bases and satisfies the
above identities.

An important
theorem proven by Drinfeld [Dr1] claims that in a
quasitriangular Hopf algebra $\G(m,\iota,S,\Delta,\varepsilon,R)$ there
exists a canonically chosen element $s\in \G$ implementing the
square of the antipode, namely $s=S(R_2)R_1$.
Its coproduct is related to the $R$-matrix by the
equation
$$\Delta(s)= (R^{op}R)^{-1}(s\o s)=(s\o
s)(R^{op}R)^{-1}\eqno(B.6)$$
which turns out to mean that $s$ defines a universal balancing
element in the category of representations of $\G$.

The universal balancing element $s$ of
$\D(H)$ takes the form
$$s:= S_{\D}(R_2)R_1\equiv \D(S^{-1}(\beta^A))\D(b_A)\eqno(B.7)$$
and is a central unitary of $\G$. Its inverse can be written
simply as
$$s^{-1}=R_1R_2=R_2R_1\,.\eqno(B.8)$$
The existence of $s$ satisfying (B.6) is needed in Section 4.1 to prove
that in the Hopf spin model
the two-point amplimorphisms (and therefore, by Lemma
3.16, {\it all} universal amplimorphisms) are strictly
translation covariant.

Summarizing, the 5-plet
$\D(H)=(\G,\Del,\varepsilon_{\D},S_{\D},R)$ defines
a quasitriangular $*$-Hopf algebra. If we compare this structure
with the double of the dual Hopf algebra $\hat H$ we find
$\D(\hat H)=(\G,\Delop,\varepsilon_{\D},S^{-1}_{\D}, R^{op})$.

\vskip 3truecm
{\eightbf References}\medskip
{\eightrm\baselineskip=9.5pt
\item{[AFS]} A. Alekseev, L. Faddeev, M. Semenov-Tian-Shansky,
Commun.Math.Phys. {\eightbf 149}, 335 (1992)
\item{[BaSk]} S. Baaj, G. Skandalis, Ann.Sci.ENS {\eightbf 26},
425 (1993)
\item{[BMT]} D. Buchholz, G. Mack, I. Todorov, Localized
automorphisms of the U(1) current algebra on the circle: An
instructive example, {\eightsl in} Algebraic theory of
superselection sectors, ed. D. Kastler, World Scientific 1990
\item{[DHR]} S. Doplicher, R. Haag and J.E. Roberts,
Commun.Math.Phys. {\eightbf 13}, 1 (1969); ibid {\eightbf 15}, 173
(1969); ibid {\eightbf 23}, 199 (1971); ibid {\eightbf 35}, 49
(1974)
\item{[DR]} S. Doplicher and J.E. Roberts, Bull.Am.Math.Soc.
{\eightbf 11}, 333 (1984);
\item{} Ann.Math. {\eightbf 130}, 75 (1989); Invent.Math.
{\eightbf 98}, 157 (1989)
\item{[Dr1]} V.G. Drinfeld, Leningrad Math. J. {\eightbf 1} 321,
(1990)
\item{[Dr2]} V.G. Drinfeld, Quantum groups, {\eightsl in} Proc.
Int. Cong. Math., Berkeley, 1986, p.798
\item{[E]} M. Enock, Kac algebras and crossed products,
Colloques Internationaux C.N.R.S., No 274 -- Algebres d'operateurs
et leurs applications en physique mathematique
\item{[ES]} M. Enock, J.M. Schwartz, "Kac algebras and duality of
locally compact groups", Springer 1992
\item{[FNW]} M. Fannes, B. Nachtergaele, R.F. Werner, Quantum spin
chains with quantum group symmetry, preprint -- KUL-TF-94/8
\item{[F]} K. Fredenhagen, Generalizations of the theory of
superselection sectors, {\eightsl in}
Algebraic Theory of Superselection Sectors, ed. D. Kastler,
World Scientific 1990
\item{[FRS]} K. Fredenhagen, K.-H. Rehren and B. Schroer,
Commun.Math.Phys. {\eightbf 125}, 201 (1989);
Rev.Math.Phys. Special Issue 113 (1992)
\item{[Fr\"o Gab]} J. Fr\"ohlich, F. Gabbiani,
Rev.Math.Phys.{\eightbf 2}, 251 (1990)
\item{[H]} R. Haag, "Local Quantum Physics", Springer 1992
\item{[J]} V.F.R. Jones, Invent.Math. {\eightbf 72}, 1 (1983)
\item{[LaRa]} R.G. Larson, D.E. Radford, J.Algebra {\eightbf 117}, 267 (1988)
\item{[LaSw]} R.G. Larson, M.E. Sweedler, Amer. J. Math. {\eightbf 91},
75 (1969)
\item{[Maj1]} S. Majid, Tannaka-Krein theorem for quasi-Hopf
algebras and other results, {\eightsl in} Contemp.Math. {\eightbf
134}, 219 (1992)
\item{[Maj2]} S. Majid, Int.J.Mod.Phys.{\eightbf A5}, 1 (1990)
\item{[MS1]} G. Mack and V. Schomerus, Commun.Math.Phys.{\eightbf
134}, 139 (1990)
\item{[MS2]} G. Mack and V. Schomerus, Nucl.Phys.{\eightbf B370},
185 (1992)
\item{[Mu]} G.J. Murphy, "C$^*$-algebras and operator theory",
Academic Press 1990
\item{[NaTa]} Y. Nakagami, M. Takesaki, ``Duality for Crossed Products
of von Neumann Algebras'', Lecture Notes in Mathematics 731, Springer 1979
\item{[N]} F. Nill, Rev.Math.Phys.{\eightbf 6}, 149 (1994)
\item{[P]} V. Pasquier, Commun.Math.Phys. {\eightbf 118}, 355
(1988); Nucl.Phys. {\eightbf B295 [FS21]}, 491 (1988)
\item{[PS]} V. Pasquier, H. Saleur, XXZ chain and quantum SU(2),
{\eightsl in} Fields, Strings, and Critical Phenomena, Les Houches
1988, eds. E. Brezin, J. Zinn-Justin, Elsevier 1989
\item{[PoWo]} P. Podl\'es, S.L. Woronowich, Commun.Math.Phys. {\eightbf 130},
381 (1990)
\item{[R1]} K.-H. Rehren, Braid Group Statistics and their
Superselection Rules,
{\eightsl in} Algebraic Theory of Superselection Sectors, ed. D.
Kastler, World Scientific 1990
\item{[R2]} K.-H. Rehren, Commun.Math.Phys.{\eightbf 145}, 123
(1992)
\item{[S]} V. Schomerus, Quantum symmetry in quantum
theory, DESY 93-018 preprint
\item{[SzV]} K. Szlach\'anyi, P. Vecserny\'es,
Commun.Math.Phys.{\eightbf 156}, 127 (1993)
\item{[Sw]} M.E. Sweedler, "Hopf algebras", Benjamin  1969
\item{[Ta]} M. Takesaki, Duality and von Neumann Algebras, {\eightsl in}
``Lectures on Operator Algebras'' eds. A. Dold and B. Eckmann, Lect. Notes
in Math. {\eightbf 247}, Springer 1972
\item{[U]} K.H. Ulbrich, Israel J.Math. {\eightbf 72}, 252 (1990)
\item{[Wa]} Y. Watatani, ``Index for C$^*$-subalgebras'', Memoires of the AMS,
No. 424 (1990)
\item{[W]} S.L. Woronovich, Commun.Math.Phys. {\eightbf 111}, 613
(1987)
 }

\vfill\eject
\end